
\input phyzzx
\def\Z{{Z}}

\let\be=\beta

\let\de=\delta

\let\ep=\varepsilon

\let\ka=\kappa
\let\la=\lambda

\let\Si=\Sigma
\let\th=\theta

\let\om=\omega
\let\Om=\Omega
\let\p=\partial
\let\<=\langle
\let\>=\rangle

\def\comment#1{ \hbox{Comment suppressed here.} }

%
\hoffset=0.2truein
\voffset=0.1truein
\hsize=6truein
\def\TITLEPAGE{\frontpagetrue}
\def\CALT#1{\hbox to\hsize{\tenpoint \baselineskip=12pt
        \hfil\vtop{\hbox{\strut PUPT-91-1288}
        \hbox{\strut CALT-68-#1}
        \hbox{\strut NSF-ITP-91-128}}}}

\def\CALTECH{
        \address{California Institute of Technology,
Pasadena, CA 91125}}
\def\TITLE#1{\vskip .5in \centerline{\fourteenpoint #1}}
\def\AUTHOR#1{\vskip .2in \centerline{#1}}
\def\ANDAUTHOR#1{\smallskip \centerline{\it and} \smallskip
\centerline{#1}}
\def\ABSTRACT#1{\vskip .2in \vfil \centerline{\twelvepoint
\bf Abstract}
        #1 \vfil}
\def\ENDTITLEPAGE{\vfil\eject\pageno=1}
\hfuzz=5pt
\tolerance=10000
\singlespace
\TITLEPAGE
\CALT{1700}
\TITLE{Quantum Field Theory of Nonabelian Strings and
Vortices}
\AUTHOR{Mark G. Alford}
\address{Institute for Theoretical Physics, University of
California,
Santa Barbara, CA 93106}
\AUTHOR{Kai--Ming Lee}
\CALTECH
\AUTHOR{John March--Russell}
\address{Department of Physics, Princeton University,
Princeton, NJ
08540}
\ANDAUTHOR{John Preskill}
\CALTECH
\ABSTRACT{We develop an operator formalism for investigating
the properties
of nonabelian cosmic strings (and vortices) in quantum field
theory.
Operators are constructed that introduce classical string
sources and
that create dynamical string loops.  The operator
construction
in lattice gauge theory is explicitly described, and
correlation functions are computed in the
strong--coupling and
weak--coupling limits.  These correlation functions
are used to study
the long--range interactions  of nonabelian strings,
taking account of charge--screening effects due to
virtual particles.
Among the phenomena investigated are
the Aharonov--Bohm interactions of strings
with charged particles,
holonomy interactions between
string loops, string entanglement, the transfer of
``Cheshire charge'' to a
string loop, and domain wall decay via spontaneous string
nucleation.
We also analyze  the Aharonov--Bohm interactions of
magnetic monopoles
with electric flux tubes in a confining gauge theory.
We propose that the
Aharonov--Bohm effect can
be invoked to distinguish among various phases of a
nonabelian gauge
theory coupled to matter.}
\ENDTITLEPAGE
\vfill\eject
\normalspace
\chapter{Introduction}

In many cases, a gauge theory that undergoes the Higgs
mechanism will contain topologically stable defects.  Among
the defects that can occur
are line defects (``cosmic strings'') in 3+1 dimensions, and
pointlike vortices in 2+1 dimensions.\Ref\lesh{See, for
example: J. Preskill,
``Vortices and Monopoles,'' in {\it Architecture
of the Fundamental Interactions at Short Distances}, ed.
P.~Ramond
and R.~Stora (North-Holland, Amsterdam, 1987).}

If the symmetry breaking pattern and the matter content of
the theory are suitable, the strings  or vortices may have
long range interactions with various particles.  These
interactions are a consequence of the Aharonov--Bohm
phenomenon---the wave function of a particle acquires a
nontrivial phase when the particle is covariantly
transported around the string.
\REFS\ab{Aharonov, Y. and Bohm, D., Phys. Rev. {\bf 119},
485 (1959).}
\REF\awr{R. Rohm, Princeton University Ph.D. Thesis
(unpublished) (1985);
M. G. Alford and F. Wilczek, Phys. Rev. Lett. {\bf 62}, 1071
(1989).}\refsend

This simple observation has interesting implications.
\REFS\kw{L. Krauss and F. Wilczek, Phys. Rev. Lett. {\bf
62},
1221 (1989).}
\REF\pk{J. Preskill and L. Krauss, Nucl. Phys. {\bf B341},
50 (1990).}
\REF\amrw{M. G. Alford, J. March-Russell and F. Wilczek,
Nucl. Phys. {\bf B337}, 695 (1990).}\refsend
Because the Aharonov--Bohm interaction has infinite range,
and no local operator can destroy an object that has an
infinite--range interaction with another object, gauge
theories with such interactions always respect nontrivial
superselection rules.  The structure of the superselection
sectors can be invoked to distinguish among the various
possible phases of a gauge theory.  Furthermore, the
existence of infinite range interactions that are
fundamentally quantum mechanical exposes the limitations of
the ``no--hair'' theorems of black hole physics.  Though a
black hole may have no {\it classical} hair, it {\it can}
carry quantum numbers that are detectable by means of the
interaction between the black hole and a cosmic
string.\REF\bowick{M. J. Bowick {\it et al.}, Phys. Rev.
Lett. {\bf 61}, 2823 (1988).}
\REF\p{J. Preskill, Phys. Scrip. {\bf T36}, 258 (1991).}
\REF\cpw{S. Coleman, J. Preskill, and F. Wilczek, Mod. Phys.
Lett. {\bf A6}, 1631 (1991);  Phys. Rev. Lett. {\bf 67},
1975 (1991)}
\refmark{\kw,\bowick-\cpw}

The ``Aharonov--Bohm phase'' acquired by an object that
circumnavigates a string is, in general, a gauge
transformation contained in the unbroken gauge group.
When the manifest  gauge symmetry is
nonabelian, this gauge transformation might not lie in the
center of the  group.  In that case, we say that the string
is ``nonabelian.''
{}~\Ref\nonab{Early discussions of nonabelian strings appear
in F. Bais,
Nucl. Phys. {\bf B170}(FS1), 32 (1980); T. W. B.
Kibble, Phys. Rep {\bf 67}, 183 (1980).}
\REF\schwarz{
A.S. Schwarz, Nucl. Phys. {\bf B208}, 141 (1982).}
\REF\cg{S. Coleman and P. Ginsparg, (1983), unpublished.}
\REF\alice{M. Alford, {\it et. al.}, Phys.~Rev.~Lett. {\bf
64}, 1632 (1990); {\bf 65}, 668 (E); Nucl.~Phys. {\bf B349},
414 (1991).}
The physical properties of a nonabelian
string are qualitatively different from the properties of an
abelian string.\refmark{\pk,\amrw,\schwarz-\alice}
In particular, a loop of nonabelian string
(or a {\it pair} of nonabelian vortices, in 2+1 dimensions)
can carry a nontrivial gauge charge, so that the loop has an
Aharonov--Bohm interaction with other strings.  Moreover,
nonabelian Aharonov--Bohm interactions involve transfer of
charge between string loops and charged particles.
Remarkably, the charge carried by a loop (which has a
topological origin) can not be localized anywhere on the
string loop or in its vicinity.  Following Ref.~\alice, we
will
refer to such unlocalized charge as ``Cheshire charge.''

If we are to appeal to the Aharonov--Bohm effect to probe
the phase structure of a gauge theory, or to investigate the
quantum physics of black holes, we must be able to discuss
interference phenomena in a language that does not rely on
weak--coupling perturbative methods;  we need a framework
that (at least in principle) takes full account of the
effects of virtual pairs and of the fluctuations of quantum
fields.
Such a framework was erected, for abelian strings (or
vortices), in Ref.~\pk.  There, operators were constructed
that
create a loop of string, or that introduce (as a classical
source) the closed world sheet of a string.  The correlation
functions of these operators can be studied to investigate
the properties of the strings, and their Aharonov--Bohm
interactions in particular.

Our main objective in this paper is to generalize the work
of Ref.~\pk\ to the case of nonabelian strings.  Because of
the
subtle and elusive physics of nonabelian strings, this
generalization is not entirely straightforward.
\Ref\newo{An earlier discussion is contained in
M. G. Alford and J. March-Russell, ``New Order
Parameters for Non-Abelian Gauge Theories,'' Princeton
University
preprint PUPT-1226 and ITP (Santa Barbara) preprint  NSF-
ITP-90-231 (1990), to
be published in Nucl. Phys. B.}
Our primary motivation comes from two considerations.
First, we seek assurance that the exotic physics of the
nonabelian Aharonov--Bohm effect, previously inferred in the
weak--coupling limit, actually survives in a fully quantum
field--theoretic treatment.  Second, we hope to construct
(nonlocal) order parameters that can be used to classify the
phases of a gauge theory.

Let us formulate the order--parameter problem more
explicitly, and in so doing, review some of the principal
results of Ref.~\pk.

A  gauge theory can have an interesting phase diagram.
Depending on its Higgs structure and on the parameters of
the Higgs
potential, the theory may be in a Coulomb (massless) phase,
a Higgs
phase, or a confinement phase.  A Higgs phase is, roughly
speaking,
characterized by the
existence of stable magnetic flux tubes, and a confinement
phase by the
existence of stable electric flux tubes.

Nonlocal gauge--invariant order parameters can be devised
that
distinguish among the various phases.  The expectation value
of the
Wilson loop operator
\Ref\w{K. Wilson, Phys. Rev. {\bf D10}, 2445 (1974).}
exhibits area--law behavior if there
are stable
electric flux tubes, and perimeter--law behavior otherwise.
The expectation value of the 't Hooft loop operator
\Ref\th{G. 't Hooft, Nucl. Phys. {\bf B138}, 1 (1978).}
exhibits area law
behavior if there are stable magnetic flux tubes, and
perimeter--law
behavior otherwise.

These order parameters are not sufficient, however, to
distinguish among
all possible phases of a general gauge theory.  Consider,
for example,
the case of an $SU(N)$ gauge theory with matter in the
fundamental
representation.  In this theory, the Wilson loop always
obeys the
perimeter law, because an electric flux tube can break via
nucleation of
a pair.
An 't Hooft loop operator can also be defined, but always
obeys
the area law.\foot{This is actually an oversimplification of
the status of the 't Hooft loop, as we will discuss in
Sections 5, 7, and 9.}  Yet the theory can have a nontrivial
phase
diagram.
Using adjoint Higgs fields, it is possible to break the
gauge group down
to its center $Z_N$.  This phase, which admits free $Z_N$
charges, is
distinguishable from the confinement phase.

\REF\frmar{K. Fredenhagen and M. Marcu, Comm. Math. Phys.
{\bf 92}, 81 (1983).}
\REF\fred{K. Fredenhagen, ``Particle Structure of Gauge
Theories, in {\it Fundamental Problems of Gauge Field
Theory}. ed. G. Velo and A. S. Wightman (Plenum, New York,
1986).}
In Ref.~\pk,
an order parameter
was described that can distinguish the free--charge phase
from the
confinement phase in an $SU(N)$ gauge theory.\foot{This
order parameter was actually discussed earlier by
Fredenhagen and Marcu in Ref.~\frmar. See also Ref.~\fred.}
The idea is
that the
free-charge phase supports stable magnetic flux tubes
(cosmic strings),
and these strings have an {\it infinite--range}
Aharonov--Bohm interaction
with $Z_N$ charges.  No such interaction can exist if $Z_N$
charges are
confined, or if $Z_N$ is spontaneously broken by a Higgs
field that
transforms as the fundamental representation.  (Indeed, the
confinement
phase and the ``Higgs'' phase with $Z_N$ completely broken
are {\it not}
distinguishable;
\Ref\fs{E. Fradkin and S. Shenker, Phys. Rev. {\bf D19},
3682 (1979); T. Banks and E. Rabinovici, Nucl. Phys. {\bf
B160}, 349 (1979); S. Dimopoulos, S. Raby, and L. Susskind,
Nucl. Phys. {\bf B173}, 208 (1980).}
this is an instance of ``complementarity.''~

To construct the order parameter, we must first devise an
operator
$F(\Sigma)$ that introduces a cosmic string world sheet on
the closed
two-dimensional surface $\Sigma$. (As we will discuss later,
this operator is closely related to the 't~Hooft loop
operator.)
Then consider the
quantity
$$
A(\Sigma,C)={ F(\Sigma)W(C) \over
\langle F(\Sigma)\rangle \langle W(C) \rangle} ~,
\eqn\A
$$
where $W(C)$ is the (fundamental representation) Wilson
loop.  In the
free-charge phase, we have
$$
{\rm lim}~\langle A(\Sigma,C)\rangle=
\exp\left( {2\pi i \over N} k(\Sigma,C)\right)~.
\eqn\lim
$$
Here the limit is taken with $\Sigma$ and $C$ increasing to
infinite
size, and with the closest approach of $\Sigma$ to $C$ also
approaching
infinity; $k(\Sigma,C)$ denotes the linking number of the
surface
$\Sigma$ and the loop $C$.  On the other hand, if there are
no free
$Z_N$ charges, then we have
$$
{\rm lim}~\langle A(\Sigma,C)\rangle=1~.
\eqn\trivial
$$
The nonanalytic behavior of $A(\Sigma,C)$ guarantees that a
well-defined
phase boundary separates the two phases.  We will refer to
$A(\Sigma,C)$ as the ``ABOP'', or ``Aharonov--Bohm Order
Parameter.''

One of our objectives in this paper is to generalize the
above
construction, and to further explore its consequences.  In
general, it
is not sufficient to determine the realization of the center
of the
gauge group, in order to distinguish all possible phases of
a gauge
theory.  For example, $SU(N)$ might break to a discrete
subgroup $H$
that is not contained in the center.  Different unbroken
groups
(including nonabelian ones) can be distinguished according
to the
varieties of cosmic stings that exist, and the nature of the
Aharonov--Bohm interactions of these strings with free
charges.  To
probe the phase structure more thoroughly, we need to
construct a more
general $F(\Sigma)$ operator.  And we will need to consider
carefully
the interpretation of the behavior of the corresponding
$A(\Sigma,C)$
operator.  The interpretation involves subtleties associated
with the
nonabelian Aharonov--Bohm effect.

\REF\frmarc{K. Fredenhagen and M. Marcu, Phys. Rev. Lett.
{\bf 56}, 223 (1986).}
In the example described above, there are other order
parameters that can serve to distinguish the free-charge
phase from the confinement
phase.\refmark{\frmar,\fred,\frmarc} But we believe that the
properties of strings and of their Aharonov--Bohm
interactions provide a more powerful method
for classifying phases in more general cases.

The remainder of this paper is organized as follows:
Section 2  reviews the formalism for describing
configurations of
many nonabelian vortices or strings.  We emphasize the
need to select
an (arbitrary) basepoint for the purpose of defining the
``magnetic
flux'' of a vortex, discuss the holonomy interactions
between
vortices, explain the origin of Cheshire charge, and note
that nonabelian strings generically become entangled when
they cross each other.

In Section 3, we discuss the properties of
domain walls
that are bounded by loops of nonabelian string, and observe
that
physically distinct strings can bound physically identical
walls.  Hence
when a wall ``decays'' by nucleating a loop of string,
several competing
``decay channels'' may be available.

Section 4 concerns Aharonov--Bohm interactions in abelian
gauge--Higgs
systems.  We explore the screening of such interactions due
to Higgs
condensation.  The observations in Section 3 concerning
domain wall
decay are invoked, in order to interpret the results.

In Section 5, we begin our analysis of the effects of
quantum
fluctuations on the nonabelian Aharonov--Bohm effect.  We
find that
nonperturbative fluctuations cause ambiguities in the
Aharonov--Bohm
``phase'' beyond those that occur in  leading semiclassical
theory.
We construct operators that create configurations of many
nonabelian
strings.  Two types of operators are considered.  One type
introduces
{\it classical} string sources on closed world sheets.  The
other type
creates (or destroys)  {\it dynamical} string loops.

\REF\acmr{M. Alford, S. Coleman
and J. March-Russell, Nucl. Phys. {\bf B351}, 735 (1991).}
A subtle aspect of the construction is that,
for both types, the many string configurations are {\it
coherent}.  This
means the following:  The ``magnetic flux'' carried by a
string can be characterized by an element $a$ of the
unbroken gauge group $H$.  If an object that transforms as
the representation $(\nu)$ of $H$ is transported about this
string, the Aharonov--Bohm phase that it acquires, averaged
over a basis for the representation, is
$$
{1\over n_{\nu}}\chi^{(\nu)}(a)~,
\eqn\ABa
$$
where $n_{\nu}$ is the dimension of $(\nu)$ and
$\chi^{(\nu)}$ is its character.\refmark{\amrw,\acmr}
Now if an $a$ string and a
$b$ string are combined {\it incoherently}, then the
averaged phase acquired by an object that traverses the two
strings in succession is
$$
{1\over n_\nu}\chi^{(\nu)}(a)~{1\over
n_\nu}\chi^{(\nu)}(b)~;
\eqn\ABb
$$
that is, it is  the product of the Aharonov--Bohm factors
associated with the two individual strings.
But if the two strings are patched together {\it
coherently}, then the averaged phase becomes
$$
{1\over n_\nu}\chi^{(\nu)}(ab)~.
\eqn\ABc
$$
This property, that the Aharonov--Bohm factor associated
with a pair of coherently combined strings is {\it not} just
the  product of the Aharonov--Bohm factors of the two
individual strings,
is a distinguishing feature of the nonabelian
Aharonov--Bohm effect, and one of the goals of our work has
been to
better
understand how this coherence is maintained when quantum
fluctuations of
fields are fully taken into account.

Correlation functions of
the operators constructed in Section 5 are used in Section 6
to
investigate Aharonov--Bohm interactions in a pure gauge
theory, at both
strong and weak coupling.  In Section 7, we use these
operators to study
the quantum--mechanical mixing between different string
states first discussed in Ref.~\acmr.

Holonomy interactions between pairs of vortices, or
pairs of
string loops, are considered in Section 8.  Correlation
functions are
used to demonstrate  holonomy scattering and  string
entanglement.  We also analyze a correlation function that
realizes the
transfer of charge from a charged source to a loop of
string, and the
subsequent detection of the (Cheshire) charge on the loop by
means of
its Aharonov--Bohm interaction with yet another loop of
string.

In Section 9, we consider nonabelian gauge--Higgs systems;
we analyze
the consequences of the Higgs mechanism for the stability of
strings,
and for the
screening of
Aharonov--Bohm interactions.  We discuss how nonlocal
order--parameters can be
used to explore the phase structure of such a system.

Section 10 concerns the properties of dynamical magnetic
monopoles, in a
confining gauge theory.  We
construct generalizations of the Wilson and 't Hooft
operators for a
theory with dynamical monopoles, and use these operators to
demonstrate
the Aharonov--Bohm interaction between monopoles and
electric flux tubes.
\REF\c{S. Coleman, ``The Magnetic Monopole Fifty Years
Later,''
in {\it The Unity of the Fundamental Interactions},
ed. A.~Zichichi (Plenum, New York, 1983).}
\REF\ss{M. Srednicki and L. Susskind, Nucl. Phys.
{\bf B179}, 239 (1981).}\refmark{\c,\ss}

The Appendix provides some additional details concerning
some of the
lattice calculations that are mentioned in the text of the
paper.

\chapter{Nonabelian vortices and strings}
\REF\b{M.~Bucher, Nucl.~Phys.
{\bf B350}, 163 (1991).}
We will briefly review the formalism\refmark{\b} for
describing
configurations of
many vortices (in two spatial dimensions) or many strings
(in three
spatial dimensions).  Our purpose is to remind the reader
of two
peculiar features of nonabelian strings.  First, there is an
ambiguity
when two or more loops of string are patched together to
construct a
multi--string configuration.  Second, a loop of string can
carry
charges, and can exchange charge with other objects by means
of the
nonabelian Aharonov--Bohm effect.  An understanding of these
features is
needed in order to interpret the behavior of the
correlation functions that we will construct.

\section{Many--Vortex Configuration}
We will consider vortices first, and discuss strings later.
A single
isolated vortex can be associated with an element of the
unbroken
gauge group $H$, according to
$$
a(C,x_0)=P\exp\left(i\int_{C,x_0}A\cdot dx\right)~.
\eqn\flux
$$
Here $C$ is a closed oriented path, far from the vortex
core, that
encloses the vortex and begins and ends at the point $x_0$.
This group
element $a(C,x_0)$
is invariant under deformations of the path $C$ that
keep $x_0$
fixed and that avoid the vortex core.
An object that transforms as the irreducible representation
$(\nu)$ of $H$ acquires the ``Aharonov--Bohm'' phase
$D^{(\nu)}(a(C,x_0))$ when covariantly transported around
the vortex.  We require $a\in H$, because the Higgs fields
that drive the symmetry breakdown must return to their
original values when so transported.  The vortices can be
topologically classified, with the topological charge taking
values in $\pi_0(H)$; that is, a vortex with ``flux'' $a\in
H$ can be smoothly deformed to another vortex with flux
$b\in H$ if and only if $a$ and $b$ lie in the same
connected component of $H$.\foot{Here, we have implicitly
defined $H$ as the unbroken subgroup of a {\it simply
connected} underlying (spontaneously broken)
gauge group.  The global topology of the underlying group is
irrelevant, however, for the purpose of classifying the
Aharonov--Bohm interactions of the vortices.}  If $H$ is a
discrete group (as we will usually assume in this paper),
then, the topological charge is specified by an element of
$H$.

Similarly, we may associate $n$ elements of $H$ with a
configuration of
$n$ vortices.  To do so, we must choose $n$ standard paths,
all
beginning and ending at the same point $x_0$, that
circumnavigate the
various vortices.\refmark{\b}

This description of the $n$-vortex configuration evidently
suffers from
some ambiguities.  First, it is not gauge invariant.
Under a gauge transformation that takes the value $h\in H$
at the point $x_0$ (and so preserves the Higgs order
parameter at that
point), the elements $a_1,a_2,\dots,a_n$ transform according
to
$$
a_i\rightarrow ha_ih^{-1}~.
\eqn\conjugacy
$$
Thus, the configurations of a single vortex are labeled by
the conjugacy
classes of $H$.  But the gauge freedom for many vortices
involves just
one overall conjugation.  This means that specifying the
positions and
classes of all vortices is not sufficient to characterize
the
multi--vortex configuration.
Single vortices can be patched together in
various ways to construct different multi--vortex states.

Another ambiguity concerns the choice of the paths $C_i$
that enclose
the vortices.  The paths that begin and end at $x_0$ and
avoid the cores
of $n$ vortices fall into homotopy classes; these classes
are the
elements of $\pi_1(M_n,x_0)$, the fundamental group of
$M_n$,
the plane with $n$ punctures.  (This group is $F_n$, the
free group with
$n$ generators.)  By assigning elements of
$H$ to each of the generators of $\pi_1(M_n,x_0)$,
we have defined a homomorphism
from $\pi_1(M_n,x_0)$ into $H$.  This homomorphism (modulo
the one overall
conjugation) characterizes the configuration
of $n$ vortices.

But this homomorphism is still ambiguous, because the $n$
generators can be
chosen in various ways.  Consider in particular the case of
two
vortices, as shown in\FIG\conjugate{(a) Standard paths
$\alpha_1$ and $\alpha_2$, based at $x_0$, that enclose
vortex $1$ and vortex $2$.  (b) Vortex 1 winds around vortex
2.  (c) Path that, during the winding of vortex $1$ around
vortex $2$, gets dragged to $\alpha_1$.  (d) Path that gets
dragged to $\alpha_2$ during the winding.} Fig.~\conjugate.
Standard paths $\alpha_1$ and $\alpha_2$ have been chosen in
Fig.~\conjugate a that wind counterclockwise around the two
vortices.
A topologically distinct choice for the path around vortex
$1$
is shown in Fig.~\conjugate c, and a distinct choice for the
path
around vortex $2$ is shown in Fig.~\conjugate d.
Suppose now that vortex $1$ winds around
vortex $2$ (in the sense defined by the path $\alpha_2$),
and returns to its original position, as in Fig.~\conjugate
b.
We may deform our
paths during the winding so that no vortex ever crosses any
path;
then each path is mapped to the same group element after the
winding as before the winding.  But after the winding, the
final
deformed path is not homotopically equivalent to the initial
path.
Therefore, the homomorphism that maps $\pi_1(M_2,x_0)$ to
the group
has changed.

Suppose, for example, that initially $\alpha_1$ is mapped to
$a_1$ and
$\alpha_2$ is mapped to $a_2$.  We wish to determine the
final values,
after the winding, of the group elements associated with the
paths
$\alpha_1$ and $\alpha_2$.  For this purpose, it is
convenient to
notice that, during the winding, the path shown in
Fig.~\conjugate c
is ``dragged'' to $\alpha_1$.  Therefore, the group element
associated
with this path before the winding is the same as the group
element
associated with $\alpha_1$ after the winding.  We see that
this path
can be expressed as
$(\alpha_1\alpha_2)\alpha_1(\alpha_1\alpha_2)^{-1}$, where
$\alpha_1\alpha_2$ denotes the path that is obtained by
traversing
$\alpha_2$ first and $\alpha_1$ second.  The initial
homomophism
maps this path to a path--ordered exponential with the value
$$
a_1'=(a_1a_2)a_1(a_1a_2)^{-1}~.
\eqn\exchAA
$$
Similarly we note that the path shown in Fig.~\conjugate d
is
dragged during the winding to $\alpha_2$.  This path is
$(\alpha_1
\alpha_2)\alpha_2(\alpha_1\alpha_2)^{-1}$, and is mapped by
the initial
homomorphism to
$$
a_2'=(a_1a_2)a_2(a_1a_2)^{-1}~.
\eqn\exchBB
$$
We conclude that the final homomorphism after the winding
maps $\alpha_1$
to $a_1'$ and maps $\alpha_2$ to $a_2'$.
Winding vortex $1$ around
vortex $2$ has changed the two-vortex state (if $a_1$ and
$a_2$ do not commute)---both group elements have
become conjugated by $a_1a_2$.

This change in the two--vortex state
is not a mere mathematical curiosity.
The physical interpretation is
that there is
a long--range interaction between non-commuting
vortices.\REF\ww{F. Wilczek and Y.--S. Wu,
Phys. Rev. Lett. {\bf 65},
13 (1990).}\refmark{\ww,\b}
We refer to this as the ``holonomy interaction.''  It is a
type of Aharonov--Bohm interaction, but we prefer not to use
that name in this context.  The holonomy interaction is
really
a classical effect, while the term ``Aharonov--Bohm'' is
best
reserved for an intrinsically quantum--mechanical
interaction.
When the number of vortices is
three or more, these
holonomy interactions can be quite complicated.

Finally, another source of ambiguity concerns the choice of
the
basepoint $x_0$.  Winding the basepoint around some of the
vortices (or
some of the vortices around the basepoint) also changes our
description
of the multi--vortex configuration.  This feature is not of
much
physical interest.  When it is convenient, we will consider
the
basepoint to be far from all vortices, so that winding of
vortices
around the basepoint need not be considered.

The gauge--invariant content of the classical
$n$-vortex configuration
can be encoded in the Wilson loop operators
$$
W^{(\nu)}(C)={1\over
n_{\nu}}\chi^{(\nu)}\left(P\exp\left(i\oint_C A\cdot
dx\right)\right)~,
\eqn\wilson
$$
where $\chi^{(\nu)}$ denotes the character, and $n_{\nu}$
the dimension,
of irreducible representation $(\nu)$.
If $C$ is a closed path around a single vortex, then
$W^{(\nu)}(C)$, for all
$(\nu)$, provides sufficient information to identify the
class to which the
vortex belongs.  Likewise, $W^{(\nu)}(C)$, for all $(\nu)$
and all $C$ in
$\pi_1(M_n)$ suffices to determine the $n$-vortex
configuration, up to
one overall conjugation.  It is not sufficient though, to
know
$W^{(\nu)}(C_i)$ for all the generators $C_i$ of
$\pi_1(M_n)$; this determines
only the class of each vortex, but not how the vortices are
patched together.  It is therefore more convenient and
efficient to fix the
gauge at the
basepoint and assign group elements to standard paths around
the
vortices than to give a fully
gauge--invariant description in terms of Wilson loops.

When the unbroken gauge group $H$ is discrete, one sometimes
says that the theory respects a ``local discrete
symmetry.''~~
\REF\kiskis{The first formulation of local
discrete symmetries in the continuum was given by
J. Kiskis, Phys. Rev. {\bf D17}, 3196 (1978), although not
in the
context of spontaneously broken gauge theories.}
\REF\banks{T. Banks, Nucl. Phys. {\bf B323}, 90 (1989).}
\REF\li{K. Li, Nucl. Phys. {\bf B361}, 437 (1991).}
\REF\dis{M. G. Alford and J. March-Russell,
``Discrete Gauge Theories,''
Princeton preprint PUPT-91-1234 (1991), to be
published in Int. J. Mod. Phys.
B.}\refmark{\kw-\amrw,\kiskis-\dis}
This terminology is used because a field that is covariantly
constant outside the vortex core is typically not single
valued---on a closed path that encloses a vortex, a
covariantly constant field is periodic only up to the action
of an element of $H$.  (And, conversely, a field that {\it
is} periodic cannot be covariantly constant.)  But the
phrase ``local discrete symmetry'' should not be
misunderstood.  Often, when we say that a symmetry is local,
we mean that all physical states are required to be
invariant under the symmetry.  That is not what is meant
here.  Isolated $H$ charges can exist, and can, in fact, be
detected at infinite range via their Aharonov--Bohm
interactions with vortices.

\section{Cheshire Charge}
Aside from their topological charges, vortices can also
carry ordinary
$H$ quantum numbers.\refmark{\pk,\cg,\alice}  These are
easiest to
discuss if we
consider a
vortex--antivortex pair, with trivial total topological
charge.  In this
configuration, the vortex is described by a group element
$a$, and the
antivortex by $a^{-1}$.  But under $H$ transformations, $a$
mixes with
other representatives of the conjugacy class to which it
belongs.

The action of $H$ on the members of a class $\alpha$ defines
a (reducible) representation which we denote as
$D^{(\alpha)}$.
In $D^{(\alpha)}$, each element of $H$ is represented by a
permutation of the class, according to
$$
D^{(\alpha)}(h):~~\ket{a}\to\ket{hah^{-1}}~,
{}~~~a\in\alpha~.
\eqn\nonabelianD
$$
This representation can
be decomposed into irreducible
representations
of $H$.  The physical interpretation is that the vortex pair
can carry
$H$ charge.   This charge is a property of
the
vortex--antivortex
composite system
and is not localized on either vortex or antivortex.  It can
be detected
by means of the Aharonov--Bohm interaction of the
composite with another vortex.

For each class $\alpha$ in $H$  there is a unique state that
can be constructed that is uncharged (transforms trivially
under $H$); it is the superposition of group eigenstates
$$
{1\over \sqrt{n_{\alpha}}}~\sum_
{a\in\alpha}\ket{a}
\eqn\nonabelianE
$$
(where $n_\alpha$ denotes the order of the class).  Let us
consider what happens to this state when an object that
transforms as the irreducible representation $(\nu)$ of $H$
passes between the two vortices.

The composite system consisting of the vortex pair and the
charged projectile has a well-defined $H$-charge;  the
composite transforms according to a representation of $H$
that could be inferred by studying the  Aharonov--Bohm
interaction of the composite with other vortices.  (It is to
ensure this that we consider a {\it pair} of vortices with
trivial total topological charge.)  Interactions between the
vortex pair and the projectile cannot change this total
$H$-charge.  From this observation, we can infer how the
charge of the vortex pair changes when a charged object
winds around one of the vortices.\Ref\bllp{M. Bucher, K.-M.
Lee, and J. Preskill, ``On Detecting
Discrete Cheshire Charge,''
Caltech preprint CALT-68-1753 (1991).}

It is convenient to imagine that the vortex pair actually
interacts with a particle--antiparticle pair, combined in
the trivial representation of $H$, so that the total $H$-
charge
of particles plus vortices is trivial.\FIG\transfer{A
particle transforming as the representation $(\nu)$ of $H$
interacts with a vortex pair.  The particle and its
antiparticle are initially prepared in an uncharged
(gauge--singlet) state.  After the particle winds around the
vortex, the particle--antiparticle pair has acquired a
nontrivial charge.}
Then, if one particle winds around one
vortex as in Fig.~\transfer, the charge transferred to
the vortex pair is opposite to the charge transferred to the
particle pair;
the vortices are in a state transforming as the
representation $(\mu)^*$,
if the particle pair transforms as $(\mu)$.

The charge--zero state of the particle--antiparticle pair
has the group--theoretic structure
$$
{1\over \sqrt{n_\nu}}~\ket{e^{(\nu) *}_i \otimes
e^{(\nu)}_i}~,
\eqn\nonabelianF
$$
(summed over $i$)
where $\{e^{(\nu)}_i\}$ denotes a basis for the vector space
on which the representation $(\nu)$ acts, and $n_\nu$ is the
dimension of that representation.
After the particle winds around a vortex with flux $a\in
H$, this state
becomes
$$
{1\over\sqrt{n_\nu}}
\ket{e^{(\nu) *}_i \otimes e^{(\nu)}_j}
D^{(\nu)}_{ji}(a)~.
\eqn\nonabelianG
$$
We may regard this state as a vector in the space on which
the representation $D^{(\nu)*}\otimes D^{(\nu)}$ acts,
expanded in terms of the natural basis
$\ket{e^{(\nu)*}_i\otimes
e^{(\nu)}_j}$.  This vector can also be expanded in terms of
the basis that block--diagonalizes the
representation, as
$$
{1\over \sqrt{n_{\nu}}}~D^{(\nu)}(a)=
\sum_{\mu,y}c_{\mu,y}v^{(\mu),y}~.
\eqn\nonabelianH
$$
Here, $v^{(\mu),y}$ is a unit vector in the irreducible
subspace acted on by the representation $(\mu),y$; the index
$y$ occurs because a given irreducible representation may
occur in $D^{(\nu)}\otimes D^{(\nu)*}$ more than once.

We may interpret
$$
p_\mu=\sum_y|c_{\mu,y}|^2
\eqn\nonabelianI
$$
as the probability
that the final state of the particle pair transforms as
$D^{(\mu)}$;  correspondingly, it is the probability that,
after interacting with the projectile, the vortex pair
transforms as $D^{(\mu)*}$ (The $|c|$'s do not depend on how
the class representative
$a$ is chosen.)  Of course, $p_\mu$ can be nonvanishing only
if $D^{(\mu)*}$ is one of the irreducible constituents of
$D^{(\alpha)}$.

Thus we see that charge is transferred to the vortex pair,
via the nonabelian Aharonov--Bohm effect, when a charged
particle winds around the vortex;\refmark{\pk,\cg,\alice}
an initially uncharged
vortex pair becomes a nontrivial superposition of charge
eigenstates.  This is ``Cheshire charge.''  The charge is a
global
property of the vortex pair, and is not localized on the
vortices.

\section{Strings}
The above description of vortices is easily translated into
appropriate
language to describe loops of string.  A multi--string
configuration can
be characterized by assigning elements of $H$ to each of a
set of
standard paths  (beginning and ending at a basepoint $x_0$)
that link the string loops.
Again, the paths that begin and end at $x_0$ and avoid the
cores of all strings fall into homotopy classes; these
classes are the elements of $\pi_1(M,x_0)$, where $M$ now
denotes $R^3$ with all string loops removed.

\FIG\stringAB{(a) Standard paths $\alpha$ and $\beta$, based
at $x_0$, associated with the $a$ string and the $b$ string.
(b)  Path that is dragged to the path $\beta$ as the loop
of $b$ string winds through the loop of $a$ string (in the
sense defined by $\alpha$).}
Thus, when many strings are present, there is an ambiguity
in the choice of the ``standard path'' that links a given
string, and a corresponding ambiguity in the assignment of
group elements to strings.  The physics associated with this
ambiguity is that there is a long--range
holonomy interaction
between string loops.

In Fig.~\stringAB a, standard paths $\alpha$ and $\beta$
have been
introduced that wind around two string loops.  Suppose that,
initially
the homomorphism maps $\alpha$ to the group element $a$, and
maps
$\beta$ to the group element $b$.  Now suppose that the $b$
string
loops winds through the $a$ loop, in the sense defined by
$\alpha$.
During the winding, the path shown in Fig.~\stringAB b is
dragged
to $\beta$.  Thus, the flux associated with this path before
the
winding is the same as the flux associated with $\beta$
after the
winding.  This path can be expressed as
$\alpha\beta\alpha^{-1}$, and
so is mapped by the initial homomorphism to the group
element $aba^{-1}$.
We conclude that the final homomorphism after the winding
maps $\beta$
to
$$
b'=aba^{-1}~,
\eqn\strexchAAA
$$
and maps $\alpha$ to $a'=a$.  When a loop of $b$ string
winds
through a loop of $a$ string, its flux becomes
conjugated---it becomes
an $aba^{-1}$ string.

For future reference, we remark that Fig.~\stringAB\ has an
alternative interpretation.  The path shown in
Fig.~\stringAB b\
will also be dragged to $\beta$ if the $a$ string winds
around the
basepoint $x_0$, without the $b$ loop passing through the
$a$ loop
at all.  Of course, the basepoint $x_0$ is completely
arbitrary,
and selected by convention, so
this process has no actual physical effect on
the $b$ loop.  Yet, the effect of parallel transport around
the
path $\beta$ will be different than before, after the $a$
loop
has lassoed the basepoint.  We will return to this point in
Section 5.

Furthermore, a string loop, like a pair of
vortices, can
carry an $H$ charge, and can exchange charge with other
charged objects
by means of the
Aharonov--Bohm effect.

\section{Branching and entanglement}
Another property of strings will be relevant to our
subsequent
analysis; namely,
that strings can {\it
branch}.  This
feature is illustrated in \FIG\compose{The branching of a
$c=ab$ string into an $a$ string and a $b$ string.  The
standard paths, based at $x_0$, associated with each string
are indicated.}
Fig.~\compose.  There a string labeled by $c=ab$ splits into
two strings
with labels $a$ and $b$.  If the gauge group is nonabelian,
then the
precise rules for composing the group
elements at a branch point depend on our conventions---these
(arbitrary) conventions  associate each string with a path
around the
string that begins and ends at the basepoint, as shown in
Fig.~\compose.

Related to the branching phenomenon is another generic
feature of
nonabelian cosmic strings---strings that are labeled by
noncommuting
group elements become {\it entangled} when they
cross.\Ref\PoeMer{V. Po\'enaru and G. Toulouse, J. Phys.
(Paris) {\bf 38}, 887 (1977); N. D. Mermin, Rev. Mod. Phys.
{\bf 51}, 591 (1979).}\FIG\entangle{(a) The standard paths
$\alpha$, $\beta$, and $\gamma$ associated with entangled
strings.  These paths are mapped by the string homomorphism
to group elements $a$, $b$, and $c$.  (b) The paths $\beta$
and $\beta'$ are both mapped to $b$, if the bridge
connecting the strings can be removed by crossing the
strings.  (c) Standard paths associated with the ``upper
halves'' of entangled strings.  (d) The paths $\beta$
and $\beta''$ are both mapped to $b$, if the bridge
can be removed by crossing the strings ``the other
way.''}  Fig.~\entangle\
illustrates that when an $a$ string and a $b$ string cross,
they become connected by a segment of $c$
string.  The
mathematics
underlying this entanglement is the very similar to that
underlying
the
holonomy interaction between two nonabelian vortices,
in two
spatial dimensions.

Let us calculate the flux $c$ carried by the string segment
that
connects the $a$ string and the $b$ string, after the
crossing.
First, we establish our conventions by choosing standard
paths $\alpha$, $\beta$, and $\gamma$ that encircle the
string, as in Fig. \entangle a.
The group element $c$ associated with $\gamma$ can be
determined if
we demand that the bridge connecting the strings can be
removed
by re-crossing the strings.  However, even once we have
fixed all
our conventions, the element $c$ is not
uniquely determined---there
are two possible choices.  This ambiguity arises because a
crossing
event involving two oriented strings has a handedness.  A
useful way to
think about the handedness of the crossing is to imagine
that the $a$ and $b$ strings are actually  large closed
loops.  When the string loops cross, they become linked.
But the
linking number can be either $+1$ or $-1$.  (That is, we may
consider
a surface bounded by one of the loops.  This surface
inherits an orientation from the loop, defined (say) by the
right-hand rule.  If
the other loop pierces the surface in the same sense as the
outward--pointing normal, the linking number is $+1$;
otherwise, $-1$.)
The two possible linking numbers correspond to the two
possible types
of crossing events.

For one of the two types of crossings (the case in which the
the loops have linking number $-1$),  the paths $\beta$ and
$\beta'$
in Fig.~\entangle b must be mapped to the same group
element.
In terms of the standard paths defined in Fig.~\entangle a,
we have $\beta'=\alpha^{-1}\gamma\beta\alpha.$
(Recall our notation---in a composition of paths, the path
on the right is traced first.)  We therefore find
$b=a^{-1}cba$, or
$$
c=aba^{-1}b^{-1}~.
\eqn\entAA
$$
The connection with
the holonomy interaction between vortices is clarified
by Fig.~\entangle c, where standard paths $\alpha\gamma^{-
1}$ and
$\gamma\beta$ are shown that wind around the ``upper
halves'' of the entangled strings.  Since
$ac^{-1}=(ab)a(ab)^{-1}$ and $cb=(ab)b(ab)^{-1}$,
the ``flux'' carried
by the upper half strings differs from the flux carried by
the lower half strings by conjugation by $ab$ (just as
winding an $a$ vortex
counterclockwise around a $b$ vortex causes the flux of
both vortices to be conjugated by $ab$).

For the other type of crossing (the case in which the loops
have linking number $+1$), the paths $\beta$ and $\beta''$
in Fig.~\entangle d must be mapped to the same group
element.
This path is $\beta''=\alpha\beta\gamma\alpha^{-1}$, so that
$b=abca^{-1}$, or
$$
c=b^{-1}a^{-1}ba~.
\eqn\entBB
$$
Now the upper half strings carry flux $(ab)^{-1}a(ab)$ and
$ab^{-1}b(ab)$, which is just the change in a two--vortex
state
that results from winding an $a$ vortex around a $b$ vortex
in
the {\it clockwise} sense.

\chapter{Nonabelian walls bounded by strings}
If a gauge theory respects a local discrete symmetry $G$,
then, as we
have seen, the theory admits cosmic strings that are
classified by the
elements of $G$.  But if the group $G$ is spontaneously
broken to a
subgroup $H$, then some of these strings become boundaries
of domain
walls.\Ref\KLS{T. W. B. Kibble, G. Lazerides, and Q. Shafi,
Phys. Rev. {\bf D26}, 435 (1982).}  Here we will consider
some of the properties of
domain walls
bounded by strings.  It is necessary to understand these
properties, in
order to interpret the behavior of our correlation
functions, and
to use them to
probe the phase structure of a model.

In this section, as before, we assume that a standard choice
of gauge
has been made at a standard basepoint.  This choice allows
us to assign
a definite group element to each cosmic string, and also
fixes the
embedding of the unbroken group $H$ in $G$.

\section{Classifying walls bounded by string}
If a discrete gauge group $G$ is spontaneously broken to a
subgroup $H$,
then the cosmic strings of the theory are classified by
elements of
$H$.  The elements of $G$ that are not in $H$ are associated
not with
isolated strings, but rather with strings that bound
segments of domain
wall.

There is not, however, a one--to--one correspondence between
elements of
$G$ that are not in $H$ and the physically distinguishable
types of
wall.  This is because a single type of wall can be bounded
by more than
one type of string.  As is illustrated in
\FIG\WBS{(a) Standard paths that are mapped by the string
homomorphism to elements $a,b\in G$ that are {\it not} in
the unbroken group $H$.  Each string is the boundary of a
domain wall.  (b) A path that is mapped to $ab^{-1}\in H$.
This path does not cross the wall.}
Fig.~\WBS, strings labeled by $a$ and
$b$ can
bound the same wall, if $h=ab^{-1}$ is in the unbroken group
$H$.
The composite of the two strings is an object that
does {\it not} end on a wall, which means that the wall that
ends on a
$a$ string must also end on a $b$ string (with conventions
chosen as
in the the figure).
Thus, no locally measurable property distinguishes
a wall that ends on an $a$ string from a wall that ends on a
$b$
string, if $ab^{-1}\in H$.  In other words, walls are
classified by
the cosets $\{gH\}$ of $H$ in $G$ (with the trivial coset
corresponding to
the trivial wall, \ie,  no wall at all).  Whether the walls
are labeled
by left or right cosets is a matter of convention.

Another way to think about this classification is in terms
of the
discontinuity of the order parameter across the domain wall.
Suppose that the Higgs field $\Phi$ that drives the
breakdown from $G$ to $H$
transforms as the representation $R$ of
$G$, so that $D^{(R)}(h)\Phi=\Phi$ for $h\in H$.   Across
the
wall labeled by $a$, the order parameter jumps from $\Phi$
to $D^{(R)}(a)\Phi$.
This
discontinuity is independent of the choice of coset
representative.

To avoid confusion, we should remark that the classification
that we
have just described is not the usual classification of
domain walls in a
spontaneously broken gauge theory.  The point is that we are
not
classifying the domain walls that are absolutely stable, as
in the
standard analysis of topological defects in gauge theories.
Instead, we
are classifying the different walls that can end on strings.

Notice that the criterion for two strings to be boundaries
of the same
wall is {\it not} the same as the criterion for two strings
to be
(locally) indistinguishable objects.  Strings labeled by $a$
and $b$
have the same structure if $a$ and $b$ are related by
$a=hbh^{-1}$ for some $h\in H$.  Two elements of $G$ can be
in the
same $H$ coset without being related in this way.
Therefore, the same
type of wall can have more than one type of boundary, in
general.

\section{Wall decay}
The walls described above, those associated with elements of
$G$ that
are
not in the unbroken group $H$, are unstable.  Such a wall
will decay
by nucleating a loop of string by quantum tunneling; the
loop then
expands, consuming the wall\refmark{\lesh}.

As we have seen, it is possible for one type of wall to end
on more than
one type of string.  Hence, the wall may be able to decay in
more than
one way.  Furthermore, if a large segment of wall bounded by
an $a$ string
decays, two different decay modes that are {\it locally}
indistinguishable may be {\it globally} distinguishable.
For example,
it may be that $b$ and $b'$ are both in the same $H$ coset
as $a$, and
also that $b'=hbh^{-1}$, for some $h\in H$.  Then the wall
can decay by
nucleating either a $b$ string or a $b'$ string, and the
strings look
the same locally.  But $ab^{-1}$ and $ab'^{-1}$ need not be
the same
element of $H$.  Then the ribbon of wall that is produced by
the
nucleation of the loop (see Fig.~\WBS), is different in the
two cases.

If a large loop of $a$ string is prepared, which bounds a
wall, then the
wall can decay in any of the available channels.  The string
with the
lowest tension is the most likely to nucleate, but all
strings that can
bound the wall nucleate at some nonvanishing rate.  Many
holes eventually appear in
the wall, and the boundaries of the holes (the strings)
expand due to
the tension in the wall.  Eventually the holes collide.

When an $a$ hole meets a $b$ hole, junctions form, and an
$ab^{-1}$
string appears that bridges the hole.\FIG\hole{(a) Two holes
appear in a domain wall, due to the spontaneous nucleation
of loops of $a$ string and $b$ string.  (b) The holes meet
and coalesce.  A segment of $ab^{-1}$ string appears that
bridges the hole.  (c)  The $b$ string has higher tension
than the $a$ string, and so pulls the string junctions
toward each other.  (d)  A loop of $ab^{-1}$ string breaks
free from the decaying wall.} (See
Fig.~\hole.)  It may
be that the $b$ string has higher tension than the $a$
string.  Then the
junctions will get pulled around the boundary of the hole.
They
annihilate, liberating a $ab^{-1}$ string from the decaying
wall system.

This scenario shows that, when the wall bounding a very
large loop of
string decays, we may regard the decay process as dominated
by the
strings on which the wall can end that have the smallest
possible
tension.  This insight will help us to interpret the
behavior of the
order parameter $A(\Sigma,C)$.  If an operator creates a
world sheet of
an $a$ string, which bounds a wall,
then the Aharonov--Bohm interaction with a charge whose
world line winds around the string world sheet will be the
same as the
Aharonov--Bohm interaction of the charge with the $ab^{-
1}\in H$ string,
if the $b$ string is the lightest one that can nucleate on
the wall.

For completeness, we note another structure that can arise
in walls
bounded by strings.  We have seen that an $ab$ string can
branch into an
$a$ string and a $b$ string.  It may be that all three
strings are
boundaries of walls.  Then a possible configuration is shown
in
\FIG\vein{A vein in a wall.} Fig.~\vein.  Here the $ab$
string is a vein
in the wall, where the type of wall bounded by the $a$
string joins the
type of wall bounded by the $b$ string.

\chapter{The abelian order parameter}
The ``abelian'' order parameter introduced in Ref.~\pk\
can be used
to probe the realization of the center of the gauge group.
Here we will
briefly describe how this order parameter is used to
distinguish the
various phases of a $Z_N$ gauge--spin system on the lattice.

The model we consider has gauge variables $U_l$ residing on
the links
(labeled by $l$) of a cubic lattice, and spin variables
$\phi_i$ residing
on the sites (labeled by $i$).  Both gauge and spin
variables take values
in
$$
Z_N\equiv \{\exp(2\pi i k/N),~k=0,1,2,\dots,N-1\}~,
\eqn\Z
$$
The (Euclidean) action of the model is
$$
S=S_{\rm gauge} + S_{\rm spin}~,
$$
where
$$
S_{\rm gauge}=-\beta
\sum_P ( U_P+ c.c.)~,
\eqn\gauge
$$
and
$$
S_{\rm spin}=-\sum_{m=1}^{N-1}\gamma_m\sum_l
\left([(\phi^{-1} U
\phi)_l]^m+c.c.\right)~.
\eqn\spin
$$
Here $U_P=\prod_{l\in P}U_l$ associates with each elementary
plaquette (labeled by $P$) the (ordered) product of the four
$U_l$'s associated with the (oriented) links of the
plaquette, and $(\phi^{-1} U\phi)_{ij}=\phi_i^{-
1}U_{ij}\phi_j$, for each pair $ij$ of nearest--neighbor
sites.
In eq.~\spin, we have introduced an independent spin
coupling constant corresponding to
each
nontrivial irreducible representation of $Z_N$.

Now we must consider how the operator $F(\Sigma)$ is to be
defined in this lattice model.\REF\mp{G. Mack and V.
Petkova, Ann. Phys.
{\bf 123}, 442 (1979).}\REF\uwg{A. Ukawa, P. Windey, and A.
Guth, Phys. Rev. {\bf D21}, 1013 (1980).}\foot{The analogous
construction of the 't Hooft loop operator on the lattice
was first discussed in Ref.~\mp\ and Ref.~\uwg.}  Recall
that inserting $F(\Sigma)$ into a
Green function
is supposed to be equivalent to introducing a classical
cosmic string
source on the world sheet $\Sigma$.  On the lattice (in 3+1
dimensions),
we consider
$\Sigma$ to be a closed surface made up of plaquettes of the
{\it dual}
lattice.  There is a set $\Sigma^*$ of plaquettes of the
original
lattice that are dual to the plaquettes of
$\Sigma$.\FIG\Sigmafig{The closed curve $\Sigma$ in 2+1
dimensions.  The dashed line is $\Sigma$, comprised of links
of the dual lattice.  The plaquettes shown are those in
$\Sigma^*$, which are dual to the links of $\Sigma$.  The
links marked by arrows are those in $\Omega^*$; they are
dual to the plaquettes in a surface $\Omega$ that is bounded
by $\Sigma$.  (We may also interpret the dashed line as a
slice through the closed surface $\Sigma$, in 3+1
dimensions.)}  (See Fig.~\Sigmafig.)  The
operator
$F_n(\Sigma)$ may be expressed as
$$
F_n(\Sigma)=\prod_{P\in\Sigma^*}\exp\left(
\beta(e^{2\pi i
n/N}U_P-U_P + c.c.)\right)~.
\eqn\F
$$
In other words,
to evaluate the path integral for a Green function with
an insertion of $F_n(\Sigma)$, we modify the plaquette
action on the plaquettes
that are dual to $\Sigma$, according to
$$
U_P\rightarrow e^{2\pi i n/N}U_P,~~~P\in\Sigma^*~.
\eqn\transform
$$
This transformation of $S_{\rm gauge}$ is equivalent to
introducing
$n$ units of $Z_N$ magnetic
flux on each of the plaquettes in $\Sigma^*$.

When the surface $\Sigma$ is large, the vacuum expectation
value of $F_n$ behaves like
$$
\VEV{F_n(\Sigma)}\sim \exp\left(
-\kappa_n^{\rm (ren)}{\rm Area}(\Sigma)\right)~.
\eqn\abelianren
$$
We may interpret $\kappa_n^{\rm (ren)}$ as the
renormalization of the tension of the string source, due to
quantum fluctuations of the charged matter fields.
Eq.~\abelianren\ also has an alternative interpretation.  We
may think of $\Sigma$ as a surface lying in a time slice,
rather than as the world sheet of a string propagating
through spacetime.  Then $F_n(\Sigma)$ is the operator
$$
F_n(\Sigma)=\exp\left({2\pi i n\over N}
Q_\Sigma\right)~.
\eqn\surface
$$
where $Q_\Sigma$ is the $Z_N$ charge contained inside
the surface $\Sigma$.  Virtual pairs of charged particles
near $\Sigma$ cause the charge $Q_\Sigma$ to fluctuate.  If
the theory has a mass gap, then the charge fluctuations near
two elements of $\Sigma$ become very weakly correlated when
the elements are distantly separated.  Thus, charge
fluctuations cause the characteristic ``area--law'' falloff
of $\VEV{F_n(\Sigma)}$ in eq.~\abelianren.

Consider the case $\beta>>1$ and $\gamma_m<<1$ (for all
$m$).
In this case, the gauge variables are highly ordered, and so
we expect
that $Z_N$ charge is not {\it confined}.  Furthermore, the
spins are
disordered, so there is no Higgs condensate to {\it screen}
$Z_N$ charge
either.  Thus, free $Z_N$ charges should exist.
We can check whether this expectation is correct
by using perturbative expansions to evaluate $\langle
A(\Sigma,C)\rangle$.

Because the gauge coupling is weak, a frustrated plaquette
(one with
$U_P\ne 1$) is very costly.  An insertion of $F_n(\Sigma)$
tends to
frustrate the plaquettes in $\Sigma^*$.  These frustrations
can
be avoided, though, if the gauge variables assume a suitable
configuration.  Choose a three-dimensional hypersurface
$\Omega$
whose
boundary is $\Sigma$.  This hypersurface consists of a set
of
cubes of the
dual lattice.  Dual to these cubes is a set $\Omega^*$ of
links of the
original lattice.  (See Fig.~\Sigmafig.)  By choosing
$$
\eqalign{&U_l=e^{2\pi i n/N},~~l\in\Omega^*~,\cr
&U_l=~~1~~~~~~,~~l\not\in\Omega^*~,}
\eqn\omegaA
$$
we can avoid frustrating any plaquettes.  (This is the
lattice analog of
a ``singular gauge transformation'' that removes the string
``singularity'' on $\Sigma$.)
With the links in the configuration eq.~\omegaA\ the
fundamental
representation Wilson loop operator
$$
W(C)=\prod_{l\in C}U_l
\eqn\abelianWil
$$
acquires the
value $\exp(2\pi
i nk/N)$, where $k$ is the linking number of $C$ and
$\Sigma$.\FIG\VEVA{A slice through a closed surface $\Sigma$
that links once with the loop $C$.  $C$ intersects the
hypersurface $\Omega$ once.}  (See Fig.~\VEVA.)

Of course, this choice of link variables changes the nearest
neighbor
interactions for each pair of spins that is joined by a link
contained
in $\Omega^*$.  But the spins are strongly coupled and
highly
disordered, so that they are nearly indifferent to this
change; it is
much more costly to frustrate a plaquette than to frustrate
a link.
Therefore, the expectation value of the ABOP
$$
A_n^{(\nu)}(\Sigma,C)={F_n(\Sigma)~[W(C)]^{\nu}\over
\VEV{F_n(\Sigma)}\VEV{[W(C)]^\nu}}
\eqn\abelianA
$$
is dominated by small
fluctuations about the ``background'' eq.~\omegaA, and we
conclude that
$\VEV{A_1^{(1)}(\Sigma,C)}$ satisfies eq.~\lim.
Thus, there is an infinite range Aharonov--Bohm
interaction, and free $Z_N$ charges exist.  (Further details
of this
analysis may be found in Ref.~\pk.)

If the gauge coupling is strong ($\beta<<1$), then there are
no free
$Z_N$ charges.  The Wilson loop introduces a $Z_N$ charge as
a classical
source, but confinement causes a pair of $Z_N$ charges to be
produced,
so that the charge is dynamically shielded.  We expect that
eq.~\trivial\ is satisfied, and this can be verified in the
small-$\beta$ expansion\refmark{\pk}.

It is also interesting to consider the case in which $\beta$
is large
and some of the spin couplings $\gamma_m$ are also large.
Then the
matter field is ordered and the matter ``condensate''
screens the
charge.

For example, suppose that
$$
\eqalign{
&\gamma_{m'}>>1~,~~~m'=m~,\cr
&\gamma_{m'}<<1~,~~~m'\ne m~.}
\eqn\coupling
$$
In effect, then, the operator $\phi^m$ condenses, and $Z_N$
is broken
down to the kernel of the representation $(m)$; this is
$Z_M$, where $M$
is the greatest common factor of $N$ and $m$.  We
anticipate, therefore, that
the operator $F_n(\Sigma)$ introduces a stable cosmic string
world sheet provided that
$nm\equiv0~(mod N)$ (so that the flux carried by the string
is in the
unbroken group $Z_M$).  Otherwise, the string introduced by
$F_n(\Sigma)$
is the boundary of a domain wall.  This wall is unstable,
and decays by
nucleating a loop of string.

{\it If} we assume that the flux of the string created by
$F_n(\Sigma)$ combines
with the flux of the nucleated string to give a trivial
total flux, then
we anticipate that the ABOP behaves as
$$
\eqalign{
&{\rm lim}~\langle
A_n^{(\nu)}(\Sigma,C)\rangle=\exp\left({2\pi i
n\nu\over
N}k(\Sigma,C)\right)~,~~mn\equiv0~({\rm mod}~N)~,\cr
&{\rm lim}~\langle
A_n(\Sigma,C)\rangle=~~~~~~~~~~1~,~~~~~~~~~~~~~~~~~~~~
mn\not\equiv0~({\rm mod}~N)~.}
\eqn\abelian
$$
{}From this behavior, we could easily infer that the unbroken
symmetry is $Z_M$;
a string has a nontrivial Aharonov--Bohm interaction with a
charge if
and only if the flux carried by the string is in $Z_M$.

However, as we emphasized in Section~3.3, even if the string
introduced by $F_n$ is the boundary of a domain wall, and
the wall decays by nucleating a string loop, it need {\it
not} be the case that the nucleated string combined with the
classical string source has trivial flux.  Thus,
eq.~\abelian\ does {\it not}, in general, correctly describe
the behavior of the ABOP in the limit
eq.~\coupling.  To see what actually happens, let us analyze
the consequences of eq.~\coupling\ using perturbative
expansions.

For $\gamma_m>>1$, spins with nontrivial $Z_M$
charge are
highly ordered, and frustrating these spins is very costly.
Now there
is a competition between the reluctance of the system to
frustrate a
plaquette (when $\beta>>1$),
and its reluctance to frustrate a $Z_M$ spin.  If the
operator $F_n(\Sigma)$ is inserted, then, as we have seen,
frustrated
plaquettes can be avoided if the links of $\Omega^*$ are
excited.  But,
if $\exp(2\pi i n/N)\not\in Z_M$,
we can excite the $U_l$'s on these links only at the cost of
frustrating
the spins there.  The number of links in $\Omega^*$
increases like the
volume enclosed by $\Sigma$, and the number of plaquettes in
$\Sigma^*$
increases only as the area of $\Sigma$.  Thus, for $\Sigma$
sufficiently
large, frustrated plaquettes are favored over frustrated
links. This is just a realization on the lattice of the
decay of a domain wall by nucleation of a loop of string,
where the frustrated spins comprise the wall, and the
frustrated plaquettes comprise the nucleated string.

But as we discussed in Section~3.3, it is sometimes possible
for
a domain
wall to decay in more than one way.  To illustrate the
possibilities, we will consider two different special cases.

\section{$Z_4\to Z_2$}
Consider, for example, a $Z_4$ model.  For
$\beta>>1$ and $\gamma_2>>1$, the $Z_4$ symmetry will break
to $Z_2$.
One finds, indeed, that $\VEV{A_2^{(\nu)}(\Sigma,C)}$
behaves as in
eq.~\abelian.

Understanding the behavior of $A_1^{(\nu)}$ and
$A_3^{(\nu)}$ involves a
subtlety.
The $n=3$ string is the anti-string of the $n=1$
string; therefore, they both have the same tension.
Furthermore,
$\exp(2\pi i/4)$ and $\exp(-2\pi i/4)$ belong to the same
coset of $Z_2$
in $Z_4$, so the $n=1$ and $n=3$ strings are both boundaries
of the same
domain wall.  When the operator $F_1(\Sigma)$ is inserted,
the resulting
domain wall can decay in two different ways.
One way, the composite string that is created has trivial
$Z_2$ flux; the other way, it has nontrivial $Z_2$ flux.

Correspondingly, when $F_1(\Sigma)$ is inserted, we may
choose either
$U_l=1$ or $U_l=-1$ on the links of $\Omega^*$; the weak
spin coupling
in the action (the $\gamma_2$ term)
depends only on $U_l^2$, and so is not frustrated either
way.  But we also need to consider the dependence on the
strong
couplings $\gamma_1$ and $\gamma_3$.  Expanding in these
small
parameters, one finds that the effective tension of the
composite string
is renormalized by spin fluctuations.  The renormalization
raises the
tension of the composite string with nontrivial $Z_2$ flux
relative to
that with trivial flux.  Thus, the configuration such that
the composite
string has trivial $Z_2$ flux really does dominate when
$F_1(\Sigma)$
(or $F_3(\Sigma)$) is
inserted.  Therefore, $A_1^{(\nu)}$ and $A_3^{(\nu)}$  do
behave as in
eq.~\abelian.

\section{$Z_6\to Z_3$}
Now consider a $Z_6$ model.  For $\beta>>1$ and
$\gamma_3>>1$, the $Z_6$ symmetry will break to $Z_3$.
$A_2^{(\nu)}$ and $A_4^{(\nu)}$ behave as in eq.~\abelian;
so do  $A_1^{(\nu)}$ and $A_5^{(\nu)}$.

Now consider $A_3^{(\nu)}$.  The $n=3$ string that is
introduced by $F_3$ is the boundary of a domain wall.  But
this wall can terminate on an $n=1$ or $n=5$ string, as well
as on an $n=3$ string.  Furthermore, the $n=1$ (or $n=5$)
string has {\it lower} tension than the $n=3$ string, so
that nucleation of this string is favored.

In other words, the configurations that dominate, when
$F_3(\Sigma)$ is inserted, have $U_l=\exp(2\pi i/3)$ (or
$U_l=\exp(-2\pi i/3)$) for $l\in\Omega^*$  By exciting the
links on $\Omega^*$, we reduce the degree of frustration of
the plaquettes of $\Sigma$.  And we do so at negligible
cost, because the $m=3$ term in $S_{\rm spin}$ cannot
distinguish $U_l=\exp(\pm 2\pi i/3)$ from $U_l=1$.

Thus, the nucleated string, combined with the classical
string source, has nontrivial flux.  The combined flux can
be either $n=2$ or $n=4$, and both occur with equal
probability.
And therefore, even though $F_3$ inserts a string that
decays, $A_3$ shows nontrivial behavior; we have
$$
{\rm lim}\VEV{A_3^{(\nu)}(\Sigma,C)}=
{1\over 2}\left(e^{2\pi i\nu/3}+e^{-2\pi i\nu/3}\right)
=\cos(2\pi\nu/3)
\eqn\decayA
$$
(if $\Sigma$ and $C$ have linking number $k=1$).

Thus, for dynamical reasons, eq.~\abelian\ is not satisfied.
Nevertheless, the behavior of $A_n^{(\nu)}$ has an
unambiguous interpretation.  For example, the $\nu$-
dependence of $\VEV{A_3^{(\nu)}}$ shows that the {\it
effective} string introduced by $F_3$ has flux that takes
values in the $Z_3$ subgroup of $Z_6$.  The only possible
interpretation is that the unbroken subgroup is $Z_3$.

Indeed, the behavior of $\VEV{A_n^{(\nu)}}$ always contains
sufficient information to unambiguously identify the
unbroken subgroup of an abelian discrete local symmetry
group.

Other order parameters have been suggested that probe the
realization of the center of the gauge
group.\refmark{\frmar,\frmarc,\fred}  We will
comment on the efficacy of these in Section~9.3.

\chapter{Nonabelian strings on the lattice}
We have seen in the previous section how the Aharonov--Bohm
effect can
be used to probe the phase structure of an abelian gauge
theory.  We now
want to extend this procedure, so that it can be used in
nonabelian
theories.  The basic strategy
will be the same as before---we will
introduce strings  and charges as classical sources, and
investigate the
dynamical response of the theory to these sources.  But the
implementation of this strategy is more delicate in the
nonabelian
case.  The main stumbling block is the problem of
introducing nonabelian
cosmic strings in a lattice gauge theory, which we address
in this
section.

We will also construct operators that create or annihilate
{\it dynamical} string loops; correlation functions of these
can be used to study the dynamical properties of strings.

\section{String calibration}
We consider a theory with (discrete) gauge group $G$.  We
are interested
in how the local $G$ symmetry is realized.  Specifically, we
wish to
identify the subgroup $H$ of $G$ that admits free charges.
($G$ quantum
numbers may be confined, or may be screened by a Higgs
condensate.)

We investigate the realization of $G$ by assembling a
laboratory that is
equipped with cosmic string loops.
As described in Ref.~\acmr\ we can
calibrate the string
loops with  a beam of particles that transform as some
faithful (not necessarily irreducible)
representation $(R)$ of $G$.  We choose an arbitrary
basepoint $x_0$,
and a basis for the representation $(R)$ at that point.
We direct the
beam from the basepoint to a beam splitter, allow the two
beams to pass
on either side of the string, and then recombine the beams
and study the
resultant interference pattern.

If the string is in a ``group eigenstate'' with flux (as
defined in eq.~\flux) $a\in G$, and $\ket{u}$ is the wave-
function
in internal--symmetry space of a particle at the basepoint,
then, when the particle is transported around a closed path
that begins and ends at $x_0$,  the wave-function is
modified according to
$$
\ket{u}\rightarrow D^{(R)}(a)~\ket{u}~.
\eqn\NAlatticeA
$$
By observing the
interference
pattern, we can measure
$$
\bra{u}D^{(R)}(a)\ket{u}~.
\eqn\NAlatticeB
$$
By varying $\ket{u}$, we can then determine all matrix
elements of
$D^{(R)}(a)$, and hence $a$ itself.  Given our choice of
basepoint, and a  choice of a basis for $D^{(R)}$, this
procedure allows us to associate a well--defined group
element with each string.

In the case where free $G$ charges may not exist, we must be
careful that the separation between the two beams remains
small compared
to any confinement distance scale or Higgs screening length.
At the
same time, of course, the separation must be large compared
to the
thickness of the string core; we assume that the core
is small compared to the
characteristic distance scale of the dynamics that we wish
to study.

Once we have calibrated the strings by measuring their
Aharonov--Bohm
interactions with nearby charges, we probe the dynamics of
the theory by
measuring the Aharonov--Bohm interactions of strings with
distant
charges.  In this way, we hope to learn what quantum numbers
are
confined or screened, and to infer the ``unbroken'' subgroup
$H$.

(As we have noted, each string loop associated with an
element of $G$ that
is not in $H$ will become the boundary of a domain wall.
Thus it might
seem that a good way to identify $H$ is to observe which
strings are
attached to walls.  Indeed, at sufficiently weak coupling,
this is a
sensible procedure, because the walls are very long lived.
But at
intermediate coupling this procedure may fail, because the
walls decay
rapidly by nucleating string loops.  The time scale for the
decay may be
comparable to the time required to assemble and calibrate a
string
loop.)

Some problems with the calibration procedure should be
pointed out.  The
first is that a string loop associated with a definite group
element is
not an eigenstate of the Hamiltonian of the theory.  An $a$
string and
a $b$ string will mix with each other if $a$ and $b$ are in
the
same $H$ conjugacy class,\refmark{\acmr} and the energy
eigenstates will be
linear
combinations of group eigenstates that transform as definite
representations of the unbroken group $H$.  (We will have
more to say
about this mixing in Section~7.2.)  But the time scale for
this mixing
increases exponentially with the size of the string loop.
For our
purposes, it will usually be legitimate to ignore the mixing
and regard
the strings as group eigenstates.

But there is another more serious problem.  While the
quantum
fluctuations that change the identity of a string loop are
very rare
(when the loop is large), there are other, much less rare,
quantum
fluctuations that can change the
Aharonov--Bohm phase that is detected.  Suppose, for
example, that a
virtual $b$ loop nucleates, lassoes the basepoint $x_0$ and
then
reannihilates.  Naturally, this process has no physical
effect on an
$a$ loop that is far from the basepoint.  Yet it changes the
Aharonov--Bohm phase acquired by a particle in
representation $(\nu)$ that
winds around
the string, beginning and ending at $x_0$; we saw in
Section 2.3 that the phase becomes $D^{(\nu)}(bab^{-1})$
rather than $D^{(\nu)}(a)$.\FIG\ambiguity{(a)  Spacetime
diagram
(in 2+1 dimensions) showing a virtual $b$ vortex--antivortex
pair that nucleates, winds around the basepoint $x_0$, and
reannihilates.  If the path based at $x_0$ at time $t_1$ is
assigned the flux $a$, then the path at time $t_2$ is
assigned flux $bab^{-1}$.
(b)  The same process, but with $x_0$ now taken to be
a fixed point in Euclidean spacetime.  The flux of the
classical $a$ vortex is measured to be $bab^{-1}$, due
to the effect of the virtual $b$ vortex pair.
Shaded areas are surfaces
bounded by the vortex world lines.}  (This
process is depicted, in 2+1 dimensions, in
Fig. \ambiguity.)  Such
fluctuations are suppressed at sufficiently weak coupling,
but they are
present, at some level, for any finite coupling.  They
result in an
ambiguity in the Aharonov--Bohm phase associated with a
string, even if
we fix the basepoint and a basis for the faithful
representation
$(\nu)$.

To avoid this ambiguity, we are forced to take a trace of
the
representation; the character $\chi^{(\nu)}(a)$ is
unaffected by these
fluctuations.  Thus, the existence of free $G$ charges (and
of an
unscreened Aharonov--Bohm interaction) can be probed by the
Wilson loop
operator eq.~\wilson.  (And since $W^{(\nu)}(C)$ is gauge
invariant,
there is no need for the loop $C$ to contain the basepoint
$x_0$.)
Let $F_a(\Sigma, x_0)$ denote the operator that
inserts a string world sheet on $\Sigma$; the string is
associated with
$a\in G$, relative to the basepoint $x_0$.  (We consider
below how this
operator is constructed.)  Then we may define the ABOP
$$
A^{(\nu)}_a(\Sigma, x_0; C)={F_a(\Sigma,
x_0)W^{(\nu)}(C)\over
\langle F_a(\Sigma, x_0)\rangle
\langle W^{(\nu)}(C) \rangle}~.
\eqn\NAA
$$
In a phase with free $G$ charges, we expect
that\refmark{\newo}
$$
{\rm lim}~\langle A^{(\nu)}_a(\Sigma, x_0; C)\rangle={1\over
n_{\nu}}
\chi^{(\nu)}(a^{k(\Sigma, C)})~,
\eqn\NAlim
$$
where $k$ denotes the linking number.  (Even this statement
requires a
qualification, for the loop $C$ must not be permitted to
come close to
retracing itself on successive passages around $\Sigma$.)

We will consider in Section~9
how $A^{(\nu)}_a(\Sigma, x_0; C)$ is expected to behave for
other
realizations of the local $G$ symmetry.

\section{Inserting string world sheets}
Now we will consider in more detail the problem of
constructing an
operator that inserts  nonabelian strings on  specified
world sheets.
The construction will be guided by the discussion in
Section~2 of the
general formalism, and by the discussion above of the
calibration
procedure.

Given a set of disjoint closed two--dimensional surfaces
$\Sigma_1, \Sigma_2 \dots$, this operator is to introduce
string world
sheets on these surfaces, where the strings are associated
with the
group elements $a_1, a_2, \dots$.  As we have emphasized,
the strings
must be referred to a common basepoint $x_0$, and each
associated group
element depends on the choice of a standard path that begins
and ends
at $x_0$ and winds around the string.  The group elements
$a_1, a_2,
\dots$ are defined  up to one overall conjugation, which
corresponds to
a gauge transformation at the basepoint $x_0$.

Furthermore, the definition of our operator must be
insensitive to the
possible confinement or screening of $G$ quantum numbers; it
must
specify the short--distance structure of the string, and
leave it up to the
dynamics of the theory whether the string can be detected at
long
range.  It must realize the calibration procedure described
above, in
which the short--range Aharonov--Bohm interactions of the
string are
determined.

\def\Flong{F_{a_1,a_2,\dots,a_n}
(\Sigma_1,\Sigma_2,\dots,\Sigma_n,x_0)}
We wish to define an operator $\Flong$ so that an insertion
of $\Flong$\ in
a Green function introduces strings on world sheets
$\Sigma_1, \Sigma_2,
\dots, \Sigma_n$.  We will first describe the operator in
heuristic terms, and
then give a more precise description in the context of
lattice gauge
theory.  Loosely speaking, an insertion of $\Flong$\ in a
Green function
imposes a restriction on the gauge field configurations that
are
included in the path integral.  Make a choice of a standard
path that
begins at $x_0$, winds around the  surface $\Sigma_1$, and
returns to
$x_0$.  Now consider a  path $P_1$ homotopic to the standard
path that runs from $x_0$ to a point
on $\Sigma_1$, traverses an infinitesimal closed loop that
encloses
$\Sigma_1$, and then retraces    itself, returning to $x_0$.
When
$\Flong$\ is inserted, the gauge field configuration is
restricted to
satisfy
$$
P~\exp\left(i\oint_{P_1,x_0}A\cdot dx\right)=a_1~,
\eqn\flux
$$
for any such $P_1$.\FIG\defF{Paths $P_1$, $P_2$, $P_3$ used
in the construction of the operator
$F_{a_1,a_2,a_3}(\Sigma_1,\Sigma_2,\Sigma_3;x_0)$.}
Similar restrictions apply for paths
$P_2, P_3,
\dots, P_n$ that wind around $\Sigma_2, \Sigma_3, \dots,
\Sigma_n$, as in Fig.~\defF.
These restrictions
on the path integral define the operator $\Flong$.

The operator so constructed is not gauge invariant.  Of
course, if it is
inserted in a Green function with gauge--invariant
operators, its
gauge--invariant part will be projected out.  Alternatively
we may
obtain an explicitly gauge--invariant operator by averaging
over gauge
transformations at the basepoint $x_0$, obtaining
$$
{1\over n_G}\sum_{g\in G}
F_{ga_1g^{-1},ga_2g^{-1},\dots,ga_ng^{-1}}
(\Sigma_1,\Sigma_2,\dots,\Sigma_n,x_0)~,
\eqn\gaugeinv
$$
where $n_G$ is the order of the group.

We may now consider the operator
$$
A^{(\nu)}_{a_1,a_2,\dots,a_n}(\Sigma_1,\Sigma_2,\dots,\Sigma
_n,x_0;C)
={\Flong W^{(\nu)}(C)\over
\langle\Flong\rangle
\langle W^{(\nu)}(C)\rangle}~.
\eqn\wow
$$
In a phase with free $G$ charge, if the loop C is homotopic
to
$P_iP_jP_k\dots$, we have
$$
{\rm lim}~\langle A^{(\nu)}_{a_1,a_2,\dots,a_n}
(\Sigma_1,\Sigma_2,\dots,\Sigma_n,x_0;C)\rangle={1\over
n_{\nu}}
\chi^{(\nu)}(a_i a_j a_k\dots)~.
\eqn\whew
$$
Thus, when a charge passes around several string loops in
succession, the Aharonov--Bohm phases acquired in each
successive passage are combined {\it coherently}.  The
coherence is maintained because we have defined the various
loops in reference to the same basepoint $x_0$.  If
different basepoints had been chosen instead, then the group
elements associated with the various string loops would have
been averaged over  conjugacy classes independently.  Since
$$
{1\over n_G}\sum_{g\in G} D^{(\nu)}(gag^{-1})={1\over
n_{\nu}}\chi^{(\nu)}
(a)~{\bf 1}
\eqn\ave
$$
(which follows from Schur's lemma), we find, for example,
$$
{\rm lim}~{\langle
F_{a_1}(\Sigma_1,x_0)F_{a_2}
(\Sigma_2,y_0)W^{(\nu)}(C)\rangle \over
\langle F_{a_1}(\Sigma_1,x_0)F_{a_2}(\Sigma_2,y_0)\rangle
\langle W^{(\nu)}(C)\rangle}={1\over
n_{\nu}}\chi^{(\nu)}(a_1)~
{1\over n_{\nu}}\chi^{(\nu)}(a_2)~,
\eqn\farout
$$
if the loop C winds around world sheets $\Sigma_1$ and
$\Sigma_2$ in succession.

If $F_{a_1}(\Sigma_1,x_0)$ and $F_{a_2}(\Sigma_2,y_0)$ have
distinct basepoints ($x_0\ne y_0$), then each by itself
introduces a gauge--singlet object.  Inserting both
operators combines two string loops  as (trivial) charge
eigenstates, rather than as group eigenstates.  This is the
reason for the lack of coherence in the Aharonov--Bohm
interaction characterized by eq.~\farout.  To combine two
strings that are not both gauge singlets, we must include in
our operator some nonlocal construct that bridges the gap
between the string loops, just as a string of electric flux
must be included in a gauge--invariant operator that creates
a widely separated quark--antiquark pair.  In the case of
two (or more) string loops, this nonlocal connection between
the loops is provided by referring the loops to a common
basepoint.

It is important, actually, that the Aharonov--Bohm
interactions of an additional loop combine incoherently with
the Aharonov--Bohm interactions of existing loops, when the
group element associated with the additional loop is
averaged over a conjugacy class.  It is this incoherence
property that ensures that the effects of {\it virtual}
string loops do not spoil the ``factorization up to a
phase'' represented by eq.~\whew.

We will now describe the construction of $F_a(\Sigma,x_0)$
in more detail, by specifying how the construction is
carried out in a lattice theory with discrete gauge group
$G$.  As discussed in Section~4, $\Sigma$ is to be regarded
as a closed surface consisting of plaquettes of the dual
lattice; these plaquettes are dual to a set $\Sigma^*$ of
plaquettes of the original lattice.  For each plaquette
$P\in\Sigma^*$, we choose a path $l_P$ on the lattice that
connects the basepoint $x_0$ to one of the corners of $P$.
These paths are chosen so that each closed loop $l_P P
l_P^{-1}$ is ``homotopic'' to the standard loop that links
$\Sigma$.
 The various paths may be chosen arbitrarily, except that
the union of all the paths should not contain any closed
loops.

The effect of the operator $F_a(\Sigma,x_0)$ is to modify
the plaquette action on each plaquette in $\Sigma^*$.
Suppose, for example, that the plaquette action is
$$
S^{(R)}_{{\rm gauge},P}= -\beta\chi^{(R)}(U_P)+
c.c.
\eqn\plq
$$
(where $R$ is a representation of $G$ that must be specified
to define the theory).
Then an insertion of $F_a(\Sigma,x_0)$
modifies the action according to
$$
S^{(R)}_{{\rm gauge},P}~\longrightarrow~
-\beta\chi^{(R)}\left(V_{l_P}a V_{l_P}^{-1}U_P\right)+
c.c.,~~~P\in \Sigma^*,
\eqn\plaqaction
$$
where
$$
V_{l_P}=\prod_{l\in l_P}U_l~.
\eqn\Vee
$$
Alternatively, we may write
$$
F_a(\Sigma,x_0)=\prod_{P\in\Sigma^*}\exp\left(\beta\chi^{(R)
}
\left(V_{l_P}a V_{l_P}^{-1}U_P\right)-\beta\chi^{(R)}(U_P)
+c.c.\right)~.
\eqn\newF
$$
The operator $\Flong$\ that inserts many string loops is
constructed by a straightforward generalization of this
procedure.

As constructed, $F_a(\Sigma,x_0)$ is not gauge invariant.
When inserted in a Green function with gauge--invariant
operators,
though,
it has the same effect as the explicitly gauge--invariant
operator
in which $a$ is averaged over a
conjugacy class, as in eq.~\gaugeinv.

It is also instructive to consider
the correlator of $F_a(\Sigma,x_0)$ with the operator
$$
U^{(\nu)}(C,x_0)=D^{(\nu)}\left(\prod_{l\in C}U_l\right)~,
\eqn\Udef
$$
where the product is taken over a closed set of links that
begins at
ends at $x_0$.  The trace of $U^{(\nu)}(C,x_0)$ is
($n_{\nu}$ times) the
Wilson loop operator $W^{(\nu)}(C)$.  But $U^{(\nu)}(C,x_0)$
itself,
like $F_a(\Sigma,x_0)$, is not invariant under a gauge
transformation
that acts nontrivially at the basepoint $x_0$.

In a phase with free G charges, and in
the leading order of weak coupling perturbation theory, one
finds
that\foot{This calculation is described in more detail in
Section~6.2 and in the Appendix.}
$$
{\rm lim}~{\langle F_a(\Sigma,x_0) U^{(\nu)}(C,x_0)\rangle
\over
\langle F_a(\Sigma,x_0)\rangle \langle{\rm
tr}~U^{(\nu)}(C,x_0)\rangle}
={1\over n_{\nu}}D^{(\nu)}\left(a^{k(\Sigma,C)}\right)~.
\eqn\untraced
$$
This equation merely states that, once a loop of string has
been
calibrated, the same Aharonov--Bohm phase can be recovered
again if
another interference experiment is subsequently performed.
But we have
already emphasized that quantum fluctuations (such as the
virtual string depicted in Fig.~\ambiguity) can spoil this
result.
Indeed, when higher orders in the weak coupling expansion
are included,
it is seen that the correlator of $F_a(\Sigma,x_0)$ and
$U^{(\nu)}(C,x_0)$ fails to ``factorize'' as in
eq.~\untraced, even when
$\Sigma$ and $C$ are far apart.  As stated before, we must
consider the
correlator of $F_a(\Sigma,x_0)$ with the gauge--invariant
operator
$W^{(\nu)}(C)$, in order to extract an Aharonov--Bohm
``phase'' that
depends only on the topological linking of $\Sigma$ and $C$
(in the
limit of infinite separation).

To conclude this section, we must ask how the operator
$F_a(\Sigma,x_0)$
depends on the basepoint $x_0$ and on the choice of the
paths $\lbrace
l_P \rbrace$.  (We consider correlation functions of $F_a$
with
gauge--invariant operators, so we regard $F_a$ as
gauge--invariant, with
$a$ averaged over a conjugacy class.)  We note that, since
the union
$\cup_P~
l_P$ of all the paths
contains no closed loops, we can choose a gauge with $U_l=1$
for
all $l\in l_P$.  In this gauge, the group element $a$ is
inserted
directly on the plaquettes of $\Sigma^*$.  Thus, it is clear
that the
correlator of $F_a(\Sigma,x_0)$
with any local gauge--invariant operator
is independent of the choice of basepoint and paths.

The correlator of $\Flong$\ with nonlocal gauge--invariant
operators (like Wilson loop operators) depends on the choice
of path only to the extent that we have already noted.  That
is, it depends on the choice of the ``standard paths'' that
enter the calibration of the string loops.\FIG\difpath{A
different choice for the path $P_2$ from the basepoint $x_0$
to the world sheet $\Sigma_2$.}  We may change
the paths from the basepoint $x_0$ to the plaquettes of
$\Sigma_2^*$, by winding these paths around $\Sigma_1$, as
in
Fig.~\difpath.  In
effect, this change alters the group element assigned to the
string world sheet on $\Sigma_2$; $a_2$ becomes replaced by
$a_1 a_2 a_1^{-1}$.  As we noted in Section~2, this
ambiguity in $\Flong$\ is the origin of the holonomy
interaction between string loops (or
vortices).\refmark{\ww,\b}

\section{Inserting string loops}
An insertion of the operator $F_a(\Sigma,x_0)$ introduces a
string on
the closed world sheet  $\Sigma$.  The string may be
regarded as
an infinitely heavy classical source.  Thus
$F_a(\Sigma,x_0)$ is closely
analogous to the Wilson loop operator $W^{(\nu)}(C)$; an
insertion of
$W^{(\nu)}(C)$ introduces an infinitely heavy classical
source (in
representation $(\nu)$) on the closed world line $C$.

But we will also find use at times for an operator that
creates (or
annihilates) a {\it dynamical} cosmic string.  Such an
operator can be
obtained by a simple modification of the construction
described
above---namely, the surface $\Sigma$ is chosen to be, rather
than a closed surface, a surface with nontrivial boundary
$C$.\FIG\defH{The paths $P_1$ and $P_2$, based at $x_0$, and
the open curves $\Sigma_1$ and $\Sigma_2$ used in the
definition of the 't Hooft operator
$B_{a,b}(x_1,y_1,\Sigma_1,x_2,y_2,\Sigma_2,x_0)$, in 2+1
dimensions.}  (See Fig.~\defH.)
Carrying out
the same procedure as before for such a surface, we arrive
at an
operator $B_a(C,\Sigma,x_0)$ that creates a string (or
annihilates an
anti-string) on the loop $C$.  More generally, an operator
$$
B_{a_1,a_2,\dots,a_n}(C_1,\Sigma_1,C_2,
\Sigma_2,\dots,C_n,\Sigma_n,x_0)
\eqn\hooft
$$
creates strings on the loops $C_1,C_2,\dots,C_n$, with all
strings
referred to a common basepoint $x_0$.

This construction generalizes a construction devised by
't~Hooft,\refmark{\th} and we
will refer to $B$ as the ``'t Hooft loop'' operator.
However, 't Hooft
considered the situation in which the strings have no
Aharonov--Bohm
interactions with other fields.  In that case, the surface
$\Sigma$ is
an invisible gauge artifact, and $B(C)$ depends on $C$
alone.  We are
interested in strings that have Aharonov--Bohm interactions,
and in that
case $\Sigma$ is not invisible.\refmark{\pk}

To better understand why the 't Hooft operator
$B_a(C,\Sigma,x_0)$ must
depend on the surface $\Sigma$ as well as on the loop $C$,
it is helpful
to think about a theory defined in 2+1 spacetime dimensions.
In that
case, the operator $B_a(x,y,\Sigma,x_0)$ creates a vortex at
$x$ and an
anti-vortex at $y$; $\Sigma$ is a path connecting $x$ and
$y$.  But no
gauge--invariant
local operator exists that creates an isolated vortex at
$x$.
Because a vortex can be detected at infinite range via the
Aharonov--Bohm effect, there is a vortex superselection
rule.  Hence,
the operator that creates a vortex cannot be local; it has a
semi-infinite string that can be seen by the fields of the
theory.
(Similarly, there is an electric charge superselection rule
in quantum
electrodynamics.  No gauge--invariant local operator can
create an
isolated electron; the operator that creates an electron
must also
create a string of electric flux that ends on the electron.)

Correlation functions such as $\langle
B_{a_1,a_2}(C_1,\Sigma_1,C_2,\Sigma_2,x_0)\rangle$ can be
used to
determine the tension of a dynamical string, or the
amplitude for mixing
between group eigenstates, as we will describe in Section~7.

(As an aside, we remark that an 't Hooft loop, like a Wilson
loop,
admits an alternative interpretation.  If a Wilson loop
operator acts on
a timelike slice, it is natural to interpret it as an
insertion of a
classical charged source, as noted above.  But if the Wilson
loop acts
on a spacelike slice, we may interpret it as an operator
that creates a
closed  electric flux tube.  We have noted that we may think
of an 't~Hooft operator acting on a spacelike
slice as an object that creates a
cosmic string.  Alternatively, we may  interpret the an 't
Hooft loop
acting on a timelike slice as an insertion of a classical
{\it magnetic
monopole} source.  In the situation originally considered by
't Hooft,
the monopole satisfied the Dirac quantization condition, and
so its
``Dirac string'' was invisible.  We are considering a
situation in which
the Dirac string is visible; the surface $\Sigma$ bounded by
$C$ is the
world sheet of this Dirac string.)

\chapter{Classical strings in the pure gauge theory}

We will now analyze the behavior of the operator
$A_a^{(\nu)}(\Sigma,x_0;C)$, using perturbative expansions.
Here we
will consider the case of a pure gauge theory with
(discrete) gauge
group $G$.  In Section~9, we will consider the effects of
introducing
matter.

\section{Strong coupling}
Although our main interest is in the physics at weak gauge
coupling, we
will make a few comments about the strong--coupling behavior
of the pure
gauge theory.

The plaquette action of the theory is taken to be
$$
S^{(R)}_{{\rm gauge},P}=-~{\beta}~\chi^{(R)}
(U_P) + c..c~,
\eqn\puregauge
$$
where $R$ is a  representation
of $G$.
For
$\beta<<1$, this theory confines sources that transform as
certain
irreducible
representations of G.  The criterion for a source to be
confined is
easiest to state in the case where $R$ is irreducible.
In that case, a source in the irreducible representation
$(\nu)$ is {\it
not} confined if and only if there are non-negative integers
$k_1$ and
$k_2$ such that
$$\eqalign{
& (R)^{k_1}\otimes (R^*)^{k_2}\supset (\nu)~,\cr
&(R)^{k_1}\otimes (R^*)^{k_2}\supset (1)~,\cr}
\eqn\strongA
$$
where $R^*$ denotes the complex conjugate of $R$, and $(1)$
denotes the
trivial representation.  The point is that if eq.~\strongA\
is
satisfied, then it is possible for a source in
representation $(\nu)$ to
be ``screened by gluons.''  In other words, the electric
flux tube that
terminates on a $(\nu)$ source can break due to glue
fluctuations.  (The
criterion eq.~\strongA\ generalizes the familiar notion that
an adjoint
representation  source is unconfined in a strongly coupled
$SU(N)$ gauge
theory.)
If $(\nu)$ {\it is} confined, then the
expectation value
of the corresponding
Wilson loop operator decays for a large loop $C$ like
$$
\langle W^{(\nu)}(C)\rangle \sim \exp\left(-
\kappa^{(\nu)}{\rm Area}(C)\right)~,
\eqn\Wildecay
$$
where ${\rm Area}(C)$ is the minimal area of a surface
bounded by
$C$, and
$\kappa^{(\nu)}$ is the tension of a flux tube that carries
electric flux in the representation $(\nu)$.  (Of course,
virtual glue
may partially screen the source; $\kappa^{(\nu)}$ is the
tension of the
lightest flux tube that can terminate on a $(\nu)$ source.)

Since the gauge variables can see a classical string with
``magnetic flux''
$a\in G$, the quantum fluctuations of the gauge variables
renormalize the
tension of the string by an amount $\kappa_a^{({\rm ren})}$.
Thus, the expectation
value of the operator $F$ decays for a large surface
$\Sigma$ like
$$
\langle F_a(\Sigma,x_0)\rangle \sim \exp\left(-
\kappa_a^{({\rm ren})}
A(\Sigma)\right)~,
\eqn\Fdecay
$$
where $A(\Sigma)$ is the area of $\Sigma$.  The calculation
of the
leading contribution to $\kappa_a^{\rm (ren)}$, for
$\beta<<1$, is
described in the Appendix.

Now consider the operator $A_a^{(\nu)}(\Sigma,x_0;C)$.
Since there are
no infinite--range Aharonov--Bohm interactions in the
confining phase of
the theory, we might expect that
$$
{\rm lim}~\langle A_a^{(\nu)}(\Sigma,x_0;C) \rangle=1~.
\eqn\trivA
$$
This is indeed found for a representation $(\nu)$ that is
not confined
(such that the electric flux tube can break).  Different
behavior is
found, however, for a representation $(\nu)$ that is
confined.  The
operator $A_a^{(\nu)}(\Sigma,x_0;C)$ can have a nontrivial
expectation
value because the electric flux tube stretched across $C$
crosses the
surface $\Sigma$ at certain points.  In the leading order of
strong--coupling perturbation theory (and assuming that no
``partial
screening'' of the source occurs), each crossing
contributes to $A$
the factor
$$
{1\over n_{\nu}}\chi^{(\nu)}(a)
\eqn\cross
$$
(or its complex conjugate, depending on the relative
orientation of the
flux tube and $\Sigma$ at the point of crossing).  Thus,
even when
$\Sigma$ and $C$ are far apart, $\langle
A_a^{(\nu)}(\Sigma,x_0;C)\rangle$ is not a purely
topological quantity
that depends only on the linking number of $\Sigma$ and $C$.
Of course, in higher orders in the strong coupling
expansion, the
behavior of $A_a^{(\nu)}$ becomes still more complicated.

Similarly, if a loop $C$ links with two different surfaces
$\Sigma_1$
and $\Sigma_2$, the operator
$A_{a_1,a_2}^{(\nu)}(\Sigma_1,\Sigma_2,x_0;C)$ acquires a
factor
$(1/n_{\nu})\chi^{(\nu)}(a_1)$ each time the electric flux
tube crosses
$\Sigma_1$ and a factor $(1/n_{\nu})\chi^{(\nu)}(a_2)$ each
time it
crosses $\Sigma_2$ (in the leading
order of the strong--coupling
expansion).  Because of confinement, the string
world sheets
combine as trivial charge eigenstates rather than group
eigenstates,
even though both are defined with respect to a common
basepoint.

\section{Weak coupling}
For $\beta>>1$ there is no confinement, and the Wilson loop
operator
exhibits perimeter law decay for any representation $(\nu)$.

The operator $F_a(\Sigma,x_0)$ decays as in eq.~\Fdecay.
The
calculation of the leading behavior of $\kappa_a^{\rm
(ren)}$, for
$\beta>>1$, is described in the
Appendix.

When the
operator $F_a(\Sigma,x_0)$ is inserted, there is a
configuration of the
gauge variables such that no plaquettes are excited.  This
configuration can be
constructed by choosing a set $\Omega$ of cubes of the dual
lattice such
that the boundary of $\Omega$ is $\Sigma$.  Each cube in
$\Omega$ is
dual to a link of the original lattice.  The configuration
with no
excited plaquettes is
$$
\eqalign{
&U_l=a~,~~~l\in\Omega^*~,\cr
&U_l=e~,~~~l\notin\Omega^*~.}
\eqn\config
$$
This configuration is unique up to a gauge transformation.
(The gauge
transformations are deformations of $\Omega$.)

Weak--coupling perturbation theory is carried out by
expanding in the
number of excited plaquettes, and in the degree of
excitation.  In the
limit $\beta\rightarrow\infty$, the configurations with the
minimal
number of excited plaquettes dominate.  By calculating the
Wilson loop
operator for the configuration with no excited plaquettes,
we verify
eq.~\NAlim\ in the weak--coupling limit.  Thus we find, as
expected,
that $G$ charges are neither confined nor screened.
Similarly, we may
verify eq.~\whew\ and eq.~\untraced\ in this limit.

When higher--order corrections in weak--coupling
perturbation theory are
computed, we find as anticipated that eq.~\NAlim\ and
eq.~\whew\
continue to hold.  But eq.~\untraced\ does not survive.
These corrections are further discussed in the Appendix.

We wish to make one other remark
here about the weak--coupling expansion,
which might help to avoid confusion.  To
calculate the weak--coupling behavior of correlation
functions that involve the operator $\Flong$, we first
construct the configuration that has  no excited plaquettes
when $\Flong$ is inserted.  This construction is a
straightforward generalization of
eq.~\config.\FIG\inequiv{Two topologically inequivalent ways
of choosing the hypersurface $\Omega_1$ that is bounded by
$\Sigma_1$.}
However, there are topologically inequivalent
ways of choosing nonintersecting surfaces $\Omega_1,\dots,
\Omega_n$, that are bounded by $\Sigma_1,\dots,\Sigma_n$ (as
in Fig~\inequiv).  Thus, one might get the impression that
there can be two (or more) {\it gauge--inequivalent}
configurations that both have no excited plaquettes.
But this is not the case.  To see why not, it is important
to keep track of the basepoint, and of the paths from the
basepoint to the loops.  If $\Omega_1$ is distorted past
$\Omega_2$, as in Fig.~\inequiv, then $\Omega_1$ crosses the
paths from the basepoint to $\Sigma_2$. To avoid exciting
any plaquettes, then, the links contained in $\Omega_2^*$
must now take the value $a_1 a_2 a_1^{-1}$.  (This, again,
is a reflection of the holonomy interaction between
string loops.)  Therefore, a Wilson loop that crosses
$\Omega_1$ first and $\Omega_2$ second, in Fig.~\inequiv
a, behaves exactly the same way as a Wilson loop that
crosses $\Omega_2$ first and $\Omega_1$ second, in
Fig.~\inequiv b.  There is a unique gauge equivalence class
of configurations with no excited plaquettes, just as there
should be.
\chapter{Dynamical  strings and vortices}
We will now consider how the 't Hooft operator can be used
to
investigate the properties of {\it dynamical} strings (in
3+1
dimensions) and vortices (in 2+1 dimensions).

\section{String tension and vortex mass}
The operator $B_{a,a^{-1}}(C_1,\Sigma_1,C_2,\Sigma_2,x_0)$
can be used
to compute the tension of a cosmic string that carries
magnetic flux
$a$ (with a caveat described below).
This operator creates an $a$ string on $C_1$  and
annihilates it
on $C_2$.  Thus, when the loops are large and far apart, we
have
$$
\eqalign{
&\langle B_{a,a^{-1}}(C_1,\Sigma_1,C_2,\Sigma_2,x_0)
\rangle\cr
&\sim \exp\left( -\kappa_a^{({\rm
ren})}(A(\Sigma_1)+A(\Sigma_2)) \right)
\exp\left( -\kappa_a^{({\rm dyn})}A(C_1,C_2)\right)~.}
\eqn\tension
$$
Here $\kappa_a^{({\rm ren})}$ is the renormalization of the
tension of a
``classical'' string source, and $\kappa_a^{({\rm dyn})}$ is
the tension
of a dynamical string; $A(C_1,C_2)$ is the area of the
minimal surface
with boundary $C_1 \cup C_2$.\FIG\twostring{Surface of
minimal area bounded by the loops $C_1$ and $C_2$ (when the
loops are close together).}  (See
Fig.~\twostring.)  If the loops $C_1$ and $C_2$
are chosen
to be far apart compared to the correlation length of the
theory, but
close together compared to the size of the loops, then the
dependence of
$\langle B_{a,a^{-1}}(C_1,\Sigma_1,C_2,\Sigma_2,x_0)
\rangle$ on the
separation between the loops determines the tension
$\kappa_a^{({\rm
dyn})}$.

Actually, the same information  can be extracted from the
behavior of
the simpler operator $B_a(C,\Sigma,x_0)$.  For a large loop
$C$, we have
$$
\langle B_a(C,\Sigma,x_0) \rangle
\sim \exp\left(-\kappa_a^{({\rm ren})}A(\Sigma)\right)
\exp\left( -\kappa_a^{({\rm dyn})}A(C)\right)~,
\eqn\tensionsimple
$$
where $A(C)$ is the area of the minimal surface bounded by
$C$.\FIG\onestring{The classical world sheet $\Sigma_{\rm
class}$ and the dynamical world sheet $\Sigma_{\rm dyn}$
associated with the 't Hooft loop operator
$B_a(C,\Sigma_{\rm class},x_0)$.}  (See Fig.~\onestring.)
Since
$A(\Sigma)$ and $A(C)$ can be varied independently,
$\kappa_a^{({\rm
dyn})}$ can be determined.  (Or, $F_a(\Sigma,x_0)$ can be
used to
measure $\kappa_a^{({\rm ren})}$.)  The calculation of
$\kappa_a^{\rm
(dyn)}$ in weak--coupling perturbation theory is described
in the
Appendix.  (In the strong--coupling limit, we have
$\kappa^{\rm
(dyn)}=0$---there are no stable magnetic flux tubes.)

Obviously, the same procedure can be used to calculate the
mass of a
vortex, in 2+1 dimensions.

The existence of a stable string (or vortex) can itself be
used to
probe the phase structure of the theory.  In a confining
phase, stable
magnetic
flux tubes do not exist; they ``melt''  due to magnetic
disorder.  If
the $G$ gauge symmetry is spontaneously broken to a subgroup
$H$, stable
$a$
strings exist only if $a\in H$.  Otherwise, an $a$ string is
the
boundary of a domain wall, which decays as described in
Section 3.3.

If no stable $a$ string exists, then $B_a(C,\Sigma,x_0)$
might not
create a stable string.  If it does not, its expectation
value will
behave, for a large loop $C$, like
$$
\langle B_a(C,\Sigma,x_0) \rangle
\sim \exp\left(-\kappa_a^{({\rm ren})}A(\Sigma)\right)
\exp\left( -m_a^{({\rm ren})} P(C)\right)~,
\eqn\perimeter
$$
where $P(C)$ is the perimeter of $C$.  It may seem, then,
that by
measuring $\VEV{B_a(C,\Sigma,x_0)}$, and determining whether
it decays
as in eq.~\tensionsimple\ or as in eq.~\perimeter, we can
find out
whether $a$ is contained in the unbroken subgroup $H$ or
not.  However,
there are subtleties.  One problem is that there may be
tradeoff between
the dependence of $\VEV{B_a(C,\Sigma,x_0)}$ on $A(\Sigma)$
and its
dependence on $A(C)$.  Eq.~\perimeter\ will apply if the
domain wall
bounded by $\Sigma$ decays by nucleating an $a^{-1}$ string
that
completely cancels the flux of the classical string source
on
$\Sigma$.  But it may be that $\kappa_a^{\rm (ren)}$ can be
reduced if
the nucleated string only partially screens the flux of the
source.  (We
saw an instance of this phenomenon in the $Z_6$ example that
was
discussed in Section 4.)  The advantage gained from reducing
$\kappa^{\rm
(ren)}$ may more than compensate for the cost of a
nonvanishing
$\kappa^{\rm (dyn)}$; then $B_a$ will decay as in
eq.~\tensionsimple,
even though $a\not\in H$.

Another complication can arise if the unbroken subgroup $H$
is not a
normal subgroup of $G$.  For then a typical $G$ conjugacy
class
contains both elements that are in $H$ and elements that are
not in $H$.
Recall that $B_a(C,\Sigma,x_0)$
is actually averaged over the $G$ conjugacy class that
contains
$a$.  One particular $H$-class contained in this $G$-class
will dominate the asymptotic behavior of
$\VEV{B_a(C,\Sigma,x_0)}$, and
whether eq.~\perimeter\ or eq.~\tensionsimple\ applies
depends on
which class dominates.

We will return to the problem of finding a suitable order
parameter, that
can be used to identify $H$, in Section~9.

\section{Mixing}
If the discrete gauge group $G$ is unbroken, then elements
of $G$ in
the same conjugacy class are associated with strings that
transform into
each other under the action of $G$.  In the classical limit,
these
``group eigenstate''
strings are degenerate energy eigenstates.  (There is also a
further
degeneracy associated with ``parity,'' which changes the
orientation, and
so transforms the $a$ string into the $a^{-1}$ string.)
Quantum
mechanically, these states mix with one another, and the
degeneracy is
lifted.  The true energy eigenstates are ``charge
eigenstates'' that
transform according to irreducible representations of $G$
(and parity).\refmark{\acmr}

This mixing can be computed using the 't Hooft operator.  We
consider a
correlation function in which an $a$ loop is created on
$C_1$ and a
$b^{-1}$ loop is annihilated on $C_2$, where $b=gag^{-1}$
for some $g\in
G$.  It is crucial that the two strings be defined with
respect to the
same basepoint $x_0$.  Otherwise, we would average $a$ and
$b$ over the
conjugacy class independently, and the correlation function
would be
dominated by the propagation of an $a$ string from $C_1$ to
$C_2$,
rather than the mixing of an $a$ string with a $b$ string.

Let $C_1$ and $C_2$ be two congruent loops, one directly
above the other
as in\FIG\mixing{Two contributions to the mixing of an $a$
string and a $b$ string.  In (a), an $ab^{-1}$ string
spontaneously nucleates, expands, meets the $a$ string, and
converts it into a $b$ string.  In (b), the $a$ string
shrinks and annihilates, then the $b$ string nucleates and
grows.}
Fig.~\mixing.  The separation between the loops
is large compared to the correlation length of the theory,
but small
compared to the size of the loops.  If the $ab^{-1}$ string
is stable, then, in the weak coupling limit, the
correlation function will be dominated by the configuration
in Fig~\mixing a. In this configuration, the world sheets of
the $a$ and $b$
strings join, and
the loop at which they join is the boundary of the world
sheet of an
$ab^{-1}$ string.  If this configuration dominates, then
$$
\eqalign{
B_{a,b}(C_1,\Sigma_1,C_2,\Sigma_2,x_0)\sim&
\exp\left( -\kappa_a^{({\rm ren})}A(\Sigma_1)
-\kappa_b^{({\rm ren})}A(\Sigma_2)\right)\cr
&\exp\left( - \kappa_{ab^{-1}}^{({\rm dyn})}
A(C_1)\right)}~.
\eqn\mix
$$
But if the $ab^{-1}$ string is unstable, and decays to a
widely separated $a$ string and $b^{-1}$ string, then the
configuration in Fig.~\mixing b will dominate.  Here,
the world sheets of
the
dynamical strings are stretched tightly across $C_1$ and
$C_2$.  If this
configuration dominates, then
$$
\eqalign{
B_{a,b}(C_1,\Sigma_1,C_2,\Sigma_2,x_0)\sim&
\exp\left( -\kappa_a^{({\rm ren})}A(\Sigma_1)
-\kappa_b^{({\rm ren})}A(\Sigma_2)\right)\cr
&\exp\left( - (\kappa_a^{({\rm dyn})}
+\kappa_b^{({\rm dyn})})A(C_1)\right)}~.
\eqn\dontmix
$$
Thus, the mixing amplitude in the weak--coupling limit is
either
$$
e^{-S_{a,b}^{\rm (mix)}}\sim
\exp\left( - (\kappa_a^{({\rm dyn})}
+\kappa_b^{({\rm dyn})})A(C_1)\right)
\eqn\dontmixagain
$$
or
$$
e^{-S_{a,b}^{\rm (mix)}}\sim
\exp\left( - \kappa_{ab^{-1}}^{({\rm dyn})}
A(C_1)\right)~,
\eqn\mixagain
$$
whichever is larger.

Actually, these are not the most general possibilities, for
the $ab^{-1}$ string may be unstable, and may prefer to
decay in some other channel.  If the $ab^{-1}$ string decays
to $n$ widely separated stable strings that carry flux
$c_1,c_2,\dots,c_n$, then we find
$$
e^{-S_{a,b}^{\rm (mix)}}\sim
\exp\left( - \left(\sum_i\kappa_{c_i}^{({\rm dyn})}
\right)
A(C_1)\right)~.
\eqn\manydecay
$$
These results were previously derived in
Ref.~\acmr.

\chapter{Interactions}
\section{Entanglement and holonomy interactions}
The Aharonov--Bohm interaction between a classical string
source and a
classical charged source was studied in Section~6.
Correlation functions can also be used to study the
holonomy
interaction between two vortices (in 2+1 dimensions), or two
strings (in
3+1 dimensions), and also the
entanglement of strings.

In 2+1 dimensions, two {\it loops} $\Sigma_1$ and $\Sigma_2$
can link,
as shown in\FIG\linked{Linked world lines $\Sigma_1$ and
$\Sigma_2$ of a classical $a$ vortex and a classical $b$
vortex.  If $a$ and $b$ do not commute, then the classical
vortices must exchange a dynamical $aba^{-1}b^{-1}$ vortex.
The exchange occurs along the path $P$, the shortest path
that connects the two world lines.  (Compare
Fig.~\entangle.)}
Fig.~\linked. Then the
quantity $\langle F_{a,b}(\Sigma_1,\Sigma_2,x_0)\rangle$
probes what
happens when one vortex source winds around another.  As we
noted in
Section~2, there is a holonomy interaction between
the vortices
if $a$ and $b$ do not commute; both are conjugated by $ab$
\refmark{\ww,\b}.
This means
that the vortex world lines cannot close (the vortex pairs
cannot
re-annihilate) unless there is an exchange of topological
quantum
numbers between the two vortices.

An $(ab)a(ab)^{-1}$ vortex becomes an $a$ vortex, and an
$(ab)
b(ab)^{-1}$ vortex becomes a $b$ vortex, if the flux
$aba^{-1}b^{-1}$ flows from the $b$ vortex to the $a$
vortex.
\foot{Here we have used different conventions to assign
group
elements to the entangled vortex world lines than the
conventions
used in Section 2.4 to assign group elements to entangled
strings
in three spatial dimensions.}
Therefore, the
quantum numbers that must be exchanged are those of an
$aba^{-1}b^{-1}$ vortex.  We thus find, in the weak coupling
limit,
$$
\eqalign{&
\langle F_{a,b}(\Sigma_1,\Sigma_2,x_0)\cr &\sim
\exp\left(-m_a^{({\rm ren})}P(\Sigma_1)
-m_b^{({\rm ren})}P(\Sigma_2)\right)
\exp\left(-m_{aba^{-1}b^{-1}}^{({\rm dyn})}
L(\Sigma_1,\Sigma_2)\right)~.}
\eqn\vorexch
$$
Here $m^{({\rm ren})}$ is the renormalization of the mass of
a classical
vortex source, $m^{({\rm dyn})}$ is the mass of a dynamical
vortex,
$P(\Sigma)$ is the perimeter of $\Sigma$, and
$L(\Sigma_1,\Sigma_2)$ is
the length of the shortest path that connects $\Sigma_1$ and
$\Sigma_2$.  (If the $aba^{-1}b^{-1}$ vortex is unstable,
then $m_{aba^{-1}b^{-1}}^{({\rm dyn})}$ is replaced by the
sum of the masses of the vortices to which it decays.)

The leading behavior of weak--coupling perturbation theory
on the
lattice is found by identifying the configurations with the
minimal
number of excited plaquettes, as we described in Section~6.
To find the leading contribution to $\langle
F_{a,b}(\Sigma_1,\Sigma_2,x_0)\rangle$,
we choose surfaces $\Omega_1$ and $\Omega_2$  that are
bounded by
$\Sigma_1$ and $\Sigma_2$ respectively, and construct a
configuration
with $U_l=a$ on links in $\Omega_1^*$ and $U_l=b$ on links
in
$\Omega_2^*$.  But where $\Omega_1$ and $\Omega_2$
intersect, there is a
string of excited plaquettes with
$$
S^{(R)}_{{\rm gauge},P}=-~{\beta}~
\chi^{(R)}(aba^{-1}b^{-1}) +
c.c.
\eqn\excstring
$$
\FIG\intersect{A plaquette (shaded)
contained in the intersection of
$\Omega_1^*$ and $\Omega_1^*$.}(See Fig.~\intersect.)  By
choosing $\Omega_1$ and $\Omega_2$ so that this string
has minimal
length, we obtain eq.~\vorexch.

(If the $aba^{-1}b^{-1}$ vortex is unstable, then the
dominant configuration in the weak--coupling limit is found
by a slightly modified procedure.  Either $\Sigma_1$ or
$\Sigma_2$, or both, becomes the boundary of several
surfaces $\Omega_1^i$, or $\Omega_2^j$.  The link
configuration is chosen so that $U_l=c_i$ on $\Omega_1^{*i}$
and $U_l=d_j$ on $\Omega_2^{*j}$, where $\prod c_i=a$ and
$\prod d_j=b$.  These surfaces intersect along several
strings that connect the two world lines; thus, the
classical vortices exchange several separated dynamical
vortices.)

In 3+1 dimensions, strings labeled by noncommuting group
elements become
entangled when they cross, as we described in Section 2.4.
String world sheets in four dimensions generically intersect
at isolated
points.  (A point of intersection is a type of
``instanton.'')  Consider
surfaces $\Sigma_1$ and $\Sigma_2$ that cross at two points,
as shown in\FIG\stringcross{World sheets $\Sigma_1$ and
$\Sigma_2$ of a classical $a$ string and classical $b$
string that intersect at two isolated points. (Only a slice
through $\Sigma_2$ is shown.)  The indicated shaded region
is the
world sheet of a dynamical $aba^{-1}b^{-1}$ string that
connects the two classical strings.} Fig.~\stringcross.
Because
the strings entangle, an $a$ world sheet on $\Sigma_1$ and a
$b$ world
sheet on $\Sigma_2$  become joined by the world sheet of an
$aba^{-1}b^{-1}$ string.  Thus, in the weak--coupling limit,
we find
$$
\eqalign{&
\langle F_{a,b}(\Sigma_1,\Sigma_2,x_0)\cr &\sim
\exp\left(-\kappa_a^{({\rm ren})}A(\Sigma_1)
-\kappa_b^{({\rm ren})}A(\Sigma_2)\right)
\exp\left(-\kappa_{aba^{-1}b^{-1}}^{({\rm dyn})}
A(\Sigma_1,\Sigma_2)\right)~,}
\eqn\stringexch
$$
where $A(\Sigma_1,\Sigma_2)$ is the area of the minimal
surface that
joins $\Sigma_1$ and $\Sigma_2$.  Again, this result is
easily verified
using weak--coupling perturbation theory on the lattice.

\FIG\torus{World sheets $\Sigma_1$ and $\Sigma_2$ of a
classical $a$ string and a classical $b$ string.  (Only a
slice through $\Sigma_2$ is shown.)  $\Sigma_1$ is a torus
that links with $\Sigma_2$.  The shaded region is the world
sheet of a dynamical $aba^{-1}b^{-1}$ string.}
It is also instructive to consider
$\VEV{F_{a,b}(\Sigma_1,\Sigma_2,x_0)}$ where the surfaces
$\Sigma_1$ and $\Sigma_2$ have the topology shown in
Fig.~\torus.  Here $\Sigma_1$ is a torus that links once
with the sphere $\Sigma_2$.  As before, $\Omega_1$ and
$\Omega_2$ unavoidably intersect along a surface of
frustrated plaquettes; we find the leading behavior of
$\VEV{F_{a,b}(\Sigma_1,\Sigma_2,x_0)}$ by choosing
$\Omega_1$ and $\Omega_2$ so that this intersection has
minimal area.  The result is again eq.~\stringexch.  But
now, if the width of the torus is small compared to the size
of the sphere, $A(\Sigma_1,\Sigma_2)$ is the area of the
minimal slice through the torus.

To interpret this result, we recall the observation in
Section~2.3, that when a loop of $a$ string winds around a
loop of $b$ string, it becomes a loop of $bab^{-1}$ string.
We may regard Fig.~\torus\ as a depiction of a process in
spacetime, in which loops of $a$ and $a^{-1}$ string are
produced, the $a$ loop winds around the $b$ string, and the
$a$ and $a^{-1}$ strings then annihilate.  But because of
the Aharonov--Bohm interaction between the strings, this
process is disallowed; an $a^{-1}$ loop cannot annihilate a
$bab^{-1}$ loop.  Therefore,
$\VEV{F_{a,b}(\Sigma_1,\Sigma_2,x_0)}$ is suppressed by the
amplitude for the $bab^{-1}$ string to become, via quantum
tunneling, an $a$ string.  Indeed, comparing with
eq.~\mixagain, we see that
$$
e^{-S_{a,bab^{-1}}^{\rm (mix)}}\sim
\exp\left(-\kappa_{aba^{-1}b^{-1}}^{({\rm dyn})}
A^{\rm (min)}(\Sigma_1)\right)
\eqn\tormix
$$
is the suppression factor in eq.~\stringexch.

\section{Cheshire charge and the Borromean rings}
As discussed in Section~2.2, it is an inevitable consequence
of the nonabelian Aharonov--Bohm phenomenon that a loop of
string (or a pair of vortices) can carry charge, and can
exchange charge with other charged
objects.\refmark{\pk,\cg,\alice}.  Let us see how
this property is reflected in the behavior of the
correlation functions of our string
operators.\refmark{\bllp}

\FIG\borromean{The Borromean rings.  $\Sigma_1$ is the world
line of an $a$ vortex, $\Sigma_2$ is the world line of a $b$
vortex, and $C$ is the world line of a charged particle that
transforms as the representation $(\nu)$.  The charged
particle transfers charge to the $a$ vortex--antivortex
pair, and the charge is subsequently detected via the
Aharonov--Bohm interaction of the pair with the $b$ vortex.}
Consider the following process (in 2+1 dimensions), depicted
in Fig.~\borromean.  First, a pair consisting of an $a$
vortex and its anti-vortex spontaneously appears; the total
charge of the pair is trivial.  Then a charged particle in
the representation $(\nu)$ winds
counterclockwise around the $a$
vortex; thus, charge is transferred to the vortex pair, as
described in Section~2.2.  Next, a $b$ vortex winds around
the (charged) vortex pair, acquiring an Aharonov--Bohm phase
that is sensitive to the charge of the pair.  Then the
$(\nu)$ particle winds clockwise around the $a$
vortex, discharging the pair.  Finally, the $a$ vortex and
anti-vortex reunite and annihilate.

If the vortices and charged particle are treated as
classical sources, this process is captured by the
correlation function
$\VEV{F_{a,b}(\Sigma_1,\Sigma_2,x_0)~W^{(\nu)}(C)}$, where
the three loops $\Sigma_1$, $\Sigma_2$ and $C$ are
configured as in Fig.~\borromean.  This is a topologically
nontrivial joining of three loops known as the ``Borromean
rings;''  no two loops are linked, yet the loops cannot be
separated without crossing.  Because this correlation
function has the spacetime interpretation described above,
we may anticipate that, when the three loops are joined but
far apart,
$\VEV{F_{a,b}(\Sigma_1,\Sigma_2,x_0)~W^{(\nu)}(C)}$ differs
from its value for large unjoined loops by a topological
multiplicative factor---the Aharonov--Bohm phase acquired by
the $b$ vortex that winds around the charged vortex pair.

\FIG\borromeanB{Nonintersecting surfaces $\Omega_1$ and
$\Omega_2$ that are bounded by the vortex world lines
$\Sigma_1$ and $\Sigma_2$.}
To calculate $A_{a,b}^{(\nu)}(\Sigma_1,\Sigma_2,x_0;C)$,
we proceed, as usual, by finding the configuration of the
link variables such that, when $F_{a,b}$ is inserted, there
are no frustrated plaquettes.  We pick surfaces $\Omega_1$,
bounded by $\Sigma_1$, and $\Omega_2$, bounded by
$\Sigma_2$, that do not intersect; such surfaces are shown
in Fig.~\borromeanB.  We then choose (up to a gauge
transformation)
$$
\eqalign{
&U_l=a~,~~~l\in \Omega_1^*~,\cr
&U_l=b~,~~~l\in \Omega_2^*~,\cr
&U_l=e~,~~~l\not\in \Omega_1^*\cup\Omega_2^*~.\cr}
\eqn\borrotriv
$$
Now we compute $W^{(\nu)}(C)$ in this configuration.  As is
clear in Fig~\borromeanB, the loop $C$ crosses first
$\Omega_1$ in a positive sense,
then $\Omega_2$ in a negative sense,
then $\Omega_1$ in a negative sense,
and finally $\Omega_2$ in a positive sense, before
closing.    The corresponding path--ordered exponential
is $ba^{-1}b^{-1}a$, and taking a trace
yields\refmark{\bllp}
$$
{\rm lim}~\VEV{A_{a,b}^{(\nu)}(\Sigma_1,\Sigma_2,x_0;C)}
\sim {1\over n_\nu}~\chi^{(\nu)}(aba^{-1}b^{-1})~.
\eqn\borrophase
$$

We should verify that the factor eq.~\borrophase\ can be
interpreted as the Aharonov--Bohm phase acquired by a $b$
vortex that winds around a vortex pair with Cheshire charge.
The interpretation is easiest if we explicitly average $a$
over the representatives of the class to which it belongs.
If this averaging is not performed, then there is a
correlation
between the choice of  the class representative that is used
to measure the charge, and the choice of the class
representative for the string that is being measured; this
correlation makes the interpretation of the measurement more
complicated.  Of course, averaging $a$ and $b$ over class
representatives independently is equivalent to defining $a$
and $b$ with reference to distinct basepoints; we have
$$
{\rm lim}~~ {\VEV{F_a(\Sigma_1,x_0)~F_b(\Sigma_2,y_0)
{}~W^{(\nu)}(C)}\over
\VEV{F_a(\Sigma_1,x_0)}~\VEV{F_b(\Sigma_2,y_0)}
{}\VEV{W^{(\nu)}(C)}}
={1\over n_G}~\sum_{g\in G}{1\over n_\nu}~
\chi^{(\nu)}(gag^{-1}bga^{-1}g^{-1}b^{-1})~.
\eqn\cheshAB
$$

We recall from the discussion in Section~2.2 that, when a
particle in representation $(\nu)$ winds around one of the
vortices of a pair that is initially uncharged, the final
state of the vortex pair is a superposition of states with
various values of the charge.  Eq.~\cheshAB\ gives the {\it
expectation value} of the Aharonov--Bohm phase acquired by a
$b$ vortex that winds around the charged pair.  If $p_\mu$
is the probability that the vortex pair transforms as the
irreducible representation $\mu$, then we have
$$
\sum_\mu p_\mu~{1\over n_\mu}\chi^{(\mu)}(b)=
{1\over n_G}~\sum_{g\in G}{1\over n_\nu}~
\chi^{(\nu)}(gag^{-1}bga^{-1}g^{-1}b^{-1})~.
\eqn\cheshinterp
$$
Indeed, eq.\cheshinterp\ can be verified directly, with
$p_\mu$ given by eq.~\nonabelianI\ and \nonabelianH.
Details will be presented elsewhere.\refmark{\bllp}

The above discussion can be applied, with hardly any
modification, to the case of charged string loops, in 3+1
dimensions.  There is an analog of the Borromean ring
configuration, in which two disjoint closed surfaces are
joined by a closed loop, although the loop is not linked
with either surface.  Eq.~\cheshAB\ holds, if $\Sigma_{1,2}$
are surfaces, and $C$ is a loop, in this configuration.

\chapter{Gauge--Higgs systems}
\section{Weak coupling}
In this section, we will consider models with Higgs fields
coupled to gauge fields.  We wish to investigate the
screening of gauge charges---and of Aharonov--Bohm
interactions---due to the Higgs mechanism.

For a gauge system with (discrete) gauge group $G$, we
introduce a Higgs field $\phi_i$, defined on sites, that
takes values in $G$.  We then include in the action of the
model the term
$$
S_{\rm Higgs}=-\sum_{\mu}\gamma_\mu\sum_l
\left(\chi^{(\mu)}\left(\left(\phi^{-1}U\phi\right)_l
\right) + c.c\right)~,
\eqn\weakA
$$
where the sum runs over all irreducible representations
$(\mu)$ of $G$.  (Compare eq.~\spin.)  Suppose that some  of
the
$\gamma_\mu$'s are large ($>>1$), while all others are small
(or zero).  Then, we can analyze the behavior of this model
using perturbative methods.

In the weak--coupling limit $\gamma_\mu\to \infty$,
$\phi^{-1}U\phi$ becomes restricted to the kernel of the
representation $(\mu)$ at each link, and so $U_P$ is also
``frozen'' to the kernel at each plaquette.  Thus, $G$
becomes ``spontaneously broken'' to the subgroup $H={\rm
Ker}(D^{(\nu)})$.  By choosing several of the $\gamma_\mu$'s
to be large, we can break $G$ to the kernel of a reducible
representation.  Indeed, breakdown to any {\it normal}
subgroup of $G$ can be achieved in this way, since every
normal subgroup is the kernel of some representation.

Of course, we know from the usual continuum weak--coupling
analysis that more general patterns of symmetry breakdown
are possible.  In that analysis, the Higgs field resides in
the vector space on which a representation of $G$ acts, and,
in principle, the unbroken subgroup could be the stability
group of any vector in this space.  In the model with
$S_{\rm Higgs}$ given by eq.~\weakA, these more general
patterns are not obtained when all of the Higgs couplings
either vanish or are very large. They might, of course be
obtained at intermediate coupling.  After suitable
block--spin transformations are performed, the effective
theory
that describes the infrared behavior of the model would be a
``continuum'' theory for which the usual analysis could
apply.

Anyway, with the breaking of $G$ to a normal subgroup $H$
implemented as described above, we can proceed to calculate
the ABOP
$A^{(\nu)}_a(\Sigma,x_0;C)$ in weak--coupling perturbation
theory, and so probe the fate of the Aharonov--Bohm
interaction in the Higgs model.  The calculation gives the
expected results.  But there is one difficulty.

The problem is that the irreducible representation $(\nu)$
of $G$ is typically a reducible representation of the
subgroup $H$.  In general, just one of the irreducible
representations
of $H$ that is contained in $(\nu)$ will dominate the
asymptotic behavior of $W^{(\nu)}(C)$ when the loop $C$ is
large.  So it is the Aharonov--Bohm interaction of this $H$
representation with an $a$ string that is probed by
$A^{(\nu)}_a$.  In fact, if $(\nu)$ contains the trivial $H$
representation, then this will dominate at weak coupling,
and $\VEV{A^{(\nu)}_a}$ will behave trivially even though
$(\nu)$ may contain other $H$ representations that {\it do}
have Aharonov--Bohm interactions with the string.

We have encountered a general problem with the
interpretation of $\VEV{A^{(\nu)}_a}$ that arises whenever
$G$ is nonabelian and is partially broken.  If
$\VEV{A^{(\nu)}_a}$ behaves nontrivially for some choice of
$a$, then we know that an unscreened charge must dominate
$W^{(\nu)}$;  by varying $a$, we can obtain information
about the representation according to which this
charge transforms.  But if $\VEV{A^{(\nu)}_a}$ behaves
trivially for all $a$, we know only that $W^{(\nu)}$ is
dominated by a screened charge.  We have no {\it a priori}
knowledge of how this charge transforms, or of how other $H$
representations contained in $(\nu)$ interact with strings.
This problem complicates the task of inferring $H$ from the
behavior of our correlation functions.

\section{String stability}
We noted in Section~$7.1$ that, if $G$ is a discrete group,
then the existence (or not) of a stable string with flux
$a\in G$ provides a criterion for determining whether $a$ is
contained in the unbroken group $H$.  It may seem, then,
that the operator $B_a(C,\Sigma,x_0)$, which creates an $a$
string, can be used to probe the realization of the $G$
gauge symmetry.  But there are two problems.

First, when inserted in a Green function
with gauge--invariant operators,
$B_a$ is actually averaged over the $G$
conjugacy class to which $a$ belongs.  This class may
contain some elements that are in the unbroken group $H$ and
some that are not, which complicates the interpretation of
$\VEV{B_a}$.

Second, even if no element of the class that contains $a$ is
in $H$, $B_a(C,\Sigma,x_0)$ may nevertheless create a stable
string.  This can happen, as we saw in Section~$4.2$, due to
the competition between the renormalization of the tension
of the classical string that propagates on $\Sigma$ and the
tension of the dynamical string whose world sheet is bounded
by $C$.

Let us ignore the second problem for the moment, and address
the first.  We first note that
the situation is relatively simple if $H$ is a normal
subgroup of $G$.
In that case, a given $G$ conjugacy class is either entirely
contained
in $H$ or is disjoint from $H$.  By studying $B_a$ for
various class
representatives, we can determine which classes are
contained in $H$,
and so reconstruct $H$.

The general case is somewhat more complicated, and $B_a$ by
itself is
not sufficient to completely determine the unbroken subgroup
$H$.
Instead, one way to proceed is the following:  If $G=
\lbrace a_1,a_2,\dots,a_{n_G}\rbrace$ is a finite group
of order $n_G$, consider
$$
\langle
B_{a_1,a_2,\dots,a_{n_G}}(C_1,\Sigma_1,C_2,\Sigma_2,\dots,
C_{n_G},\Sigma_{n_G},x_0) \rangle~,
\eqn\BigB
$$
in which all group elements appear.  Choose the loops to be
large and
widely separated, with their sizes ordered according to
$$
A(C_1)>>A(C_2)>>\cdots>>A(C_{n_G})~.
\eqn\sizes
$$
Now $B_{a_1,a_2,\dots}$ is effectively averaged over gauge
transformations at the basepoint, and so may be replaced by
$$
{1\over n_G}\sum_{g\in G}
B_{ga_1g^{-1},ga_2g^{-1},\dots,ga_{n_G}g^{-1}}
(C_1,\Sigma_1,C_2,\Sigma_2,\dots,
C_{n_G},\Sigma_{n_G},x_0)~.
\eqn\aveB
$$
The expectation value is dominated by the configurations
that
minimize
$$\kappa_{ga_1g^{-1}}^{({\rm ren})}A(\Sigma_1)
+\kappa_{ga_1g^{-1}}^{({\rm dyn})}A(C_1)
\eqn\minimize
$$
(where $\kappa_a^{({\rm dyn})}=0$ for $a\notin H$).
This condition may not
determine $g$ uniquely.  Among those $g$ that minimize
eq.~\minimize, the
dominant  configurations are such that
$\kappa_{a_2}^{({\rm ren})}A(\Sigma_2)
+\kappa_{a_{2}}^{({\rm dyn})}A(C_{2})$ is also minimized.
And so on.
Now by varying $A(C)$ and $A(\Sigma)$ independently, we find
the group
elements $a$ for which $\kappa_{gag^{-1}}^{({\rm dyn})}=0$.
We thus determine
the unbroken subgroup $H$ up to one overall conjugacy
$H\rightarrow gHg^{-1}$.  This ambiguity is expected; it
corresponds to
the freedom to change the embedding of $H$ in $G$ by
performing a gauge
transformation.  (Of course, the calculation of eq.~\BigB\
involves an average
with respect to this embedding.)

In passing, we have found the tension of all of the stable
strings
associated with the various elements of $H$.

If $G$ is abelian, as in the discussion in Section~$4.2$,
then we can overcome the second problem (that $B_a$ may
create a stable string even for $a\not\in G$) easily enough.
By measuring $A_a^{(\nu)}$, we can determine the
Aharonov--Bohm interactions of the string created by $B_a$,
and so
identify the flux of the string as belonging to $H$.  If $G$
is nonabelian, though, life is more complicated.  For as we
noted above, we have no {\it a priori} knowledge of what $H$
representation dominates the asymptotic decay of
$W^{(\nu)}$.  While it seems altogether physically
reasonable that the stability and Aharonov--Bohm
interactions of strings can be used to identify an unbroken
gauge group $H$, it is not so easily to specify how this
should be done with gauge--invariant correlation functions.

\section{The VOOP}
Another promising probe of charge screening in a gauge
theory is the ``vacuum overlap order parameter'' (VOOP)
proposed by Marcu and
Fredenhagen.\refmark{\frmar,\frmarc,\fred}  Let us compare
and
contrast the VOOP with the Aharonov--Bohm order parameter
that has been discussed in this paper.

Suppose that an $H$ gauge theory contains a matter field
$\Phi^{(\mu)}$ that transforms as the irreducible
representation $(\mu)$ of $H$.  If the local $H$ symmetry is
unbroken, and the representation $(\mu)$ is not confined or
screened, then (loosely speaking) the field $\Phi^\dagger$
should
create a stable particle.  We can express this in
gauge--invariant language.  If $x$ and $y$ are distantly
separated
points, and $P_{x,y}$ is a path from $y$ to $x$,
consider the nonlocal gauge--invariant operator
$$
K^{(\mu)}(x,y,P_{x,y})=\Phi^{(\mu)\dagger}_x
D^{(\mu)}\left(\left(\prod_{l\in P_{x,y}}U_l\right)\right)
\Phi^{(\mu)}_y~.
\eqn\voopA
$$
If $\Phi^{(\mu)\dagger}$ creates a free charge, then this
charge
must propagate between $x$ and $y$, as in\FIG\chargeprop{The
path $P$ from $y$ to $x$ that is used to construct the
gauge--invariant correlation function $K^{(\mu)}(x,y,P)$.  A
classical charged particle propagates along $P$, and a
dynamical charged particle propagates from $x$ to $y$.}
Fig.~\chargeprop.  Thus, we have
$$
\VEV{K^{(\mu)}(x,y,P)}\sim
\exp\left(-M_{\rm ren}^{(\mu)}~L(P)\right)
\exp\left(-M_{\rm dyn}^{(\mu)}~|x-y|\right)~,
\eqn\voopB
$$
where $M_{\rm dyn}^{(\mu)}$ is the mass of the stable
particle created by $\Phi^{(\mu)\dagger}$, and $M_{\rm
ren}^{(\mu)}$ is the renormalization of the mass of the
classical source propagating along $P$.
(Here, $L(P)$ is the length of $P$, and $|x-y|$ is the
distance from $x$ to $y$.)  Since $M_{\rm ren}^{(\mu)}$ can
be determined  independently by measuring
$\VEV{W^{(\mu)}(C)}$ (or by varying $L(P)$ with $|x-y|$
fixed), eq.~\voopB\ can be used to find $M_{\rm
dyn}^{(\mu)}$.  But if the representation $(\mu)$ is
screened (or confined), then $\VEV{K^{(\mu)}(x,y,P)}$ will
become independent of $|x-y|$ for large separation; in
effect, we have $M_{\rm dyn}^{(\mu)}=0$, with $M_{\rm
dyn}^{(\mu)}$ defined by eq.~\voopB.  Thus, Marcu and
Fredenhagen suggested that $M_{\rm dyn}^{(\mu)}>0$
if and only if $\Phi^{(\mu)\dagger}$ creates a free charge.

This construction is obviously closely similar to our
discussion in Section~$7.1$ of string stability, and it
suffers from related problems.
Suppose that, in a model with gauge group $G$, spontaneous
breakdown to a subgroup $H$ occurs.  Then, an irreducible
representation $(\mu)$ of $G$ contains various irreducible
representations of $H$.  It may be that among these
representations are some that can exist as free charges, and
others that are screened.  We might expect, then, that the
screened charges dominate $\VEV{K^{(\mu)}}$, so that $M_{\rm
dyn}^{(\mu)}=0$ even though $(\mu)$ contains some unscreened
representations of $H$. Even this is not clear;  because of
the competition between $M_{\rm dyn}$ and $M_{\rm ren}$ in
eq.~\voopB, the free charges may actually dominate.

Because of these difficulties, it is not at all easy, in
general, to identify the unbroken gauge group $H$ based on
the behavior of $\VEV{K^{(\mu)}(x,y,P)}$.

\section{Order parameters:  Some concluding remarks}
The existence of stable cosmic strings (or vortices), and of
Aharonov--Bohm interactions between strings and free
charges, can be used to identify and classify the various
phases of a gauge theory.  Yet, because of the problems
discussed above, it proves difficult to formulate a general
procedure that unambiguously specifies the realization of
the gauge symmetry, \eg, the ``unbroken'' subgroup.  This is
surprising (to us), but we are reluctant to attach any
fundamental significance to it.

In fact, though, just because of these problems, the phase
structure of certain gauge theories may be richer than one
might naively expect.  For example, if the gauge group $G$
is ``spontaneously broken'' to $H$, then, as we have
remarked, just one of the irreducible $H$ representations
contained in the $G$ representation $(\nu)$ will dominate
the asymptotic behavior of $\VEV{W^{(\nu)}(C)}$. On some
surface in the parameter space of the theory, a
``crossover'' may occur, where this $H$ representation
changes.  Thus, the order parameter $A^{(\nu)}_a$ might be
nonanalytic on this surface---the surface would be a sort of
phase boundary, even though no change in symmetry occurs
there.

For most of the discussion in this paper, we have taken the
gauge group $G$ to be a discrete group.  There is no
obstacle, however, to generalizing our results to the case
where $G$ is a continuous group.

\chapter{Electric flux tubes and dynamical monopoles}

In this paper, we have systematically discussed the
Aharonov--Bohm
interactions between magnetic flux tubes and electric
charges, which can
occur in the Higgs phase of a gauge theory.  Central to the
discussion
has been the effect of quantum--mechanical electric charge
fluctuations
on the interaction.

There is another type of Aharonov--Bohm interaction, which
can occur in
the {\it confining} phase of a gauge theory---the
interaction between an
{\it electric} flux tube and a {\it magnetic} charge.  The
existence of
such interactions has been noted previously,
\refmark{\c,\ss,\pk,\amrw} as have the
implications
concerning the ``magnetic hair'' carried by black
holes.\refmark{\pk,\cpw}
But the
effects of quantum--mechanical magnetic charge fluctuations
on this
incarnation of the Aharonov--Bohm phenomenon have not been
analyzed
before.

In this section, we will develop a quantum field--theoretic
treatment of
electric flux tubes in a confining gauge theory that
contains {\it dynamical}
magnetic monopoles, and will investigate the interactions of
flux tubes with
monopoles.  This treatment, of course, will be closely
similar to our
theory of magnetic flux tubes, in a Higgs phase.

First, we will describe (following closely the work of
Srednicki and
Susskind\refmark{\ss}) how dynamical magnetic monopoles are
introduced
into a lattice
gauge theory.  Then we will construct a generalized Wilson
loop operator
that creates an electric flux tube that has Aharonov--Bohm
interactions
with the monopoles.  By studying the properties of the
correlation functions of this
operator, we will investigate the effect of magnetic charge
fluctuations
on the Aharonov--Bohm effect.

\section{Magnetic Monopoles on the Lattice}

For definiteness, we will consider a gauge theory with gauge
group
$SU(N)$.
A pure $SU(N)$ gauge theory (no matter), or a theory with
the matter fields
transforming trivially under the center $Z_N$ of $SU(N)$,
admits
magnetic monopoles with $Z_N$ magnetic charges.  (Fields
that transform
under $Z_N$ would be able to see the Dirac string of such a
monopole;
that is, the monopole with minimal $Z_N$ charge would not
satisfy the
Dirac quantization condition.)  If $SU(N)/Z_N$ is actually
the unbroken
gauge symmetry of an underlying  theory with simply
connected
gauge group $G\supset
SU(N)/Z_N$, where $G$ is broken via
the Higgs mechanism, then such monopoles  arise
as topological solitons.  For example, there are models with
$G=Spin(N^2-1)$ broken to $ SU(N)/Z_N$ that contain $Z_N$
monopoles.

The (usual) 't Hooft loop operator
$B(C)$ inserts a world line of a $Z_N$ monopole along the
closed loop
$C$.  But the monopole introduced by an 't Hooft loop is a
classical
source, not a dynamical object.  We wish to elevate the
monopoles to the
status of dynamical degrees of freedom, and introduce a
coupling
constant that controls the effects of virtual monopoles.

It is most convenient to choose the dynamical variable to be
a ``Dirac
string field'' that is summed over in the path integral.  On
the
lattice, this field associates with each plaquette $P$ a
quantity
$\eta_P\in Z_N$, which may be regarded as the $Z_N$ magnetic
flux carried by a
Dirac string that pierces that plaquette.  If the total
magnetic flux
entering a cube of the lattice is nontrivial, then a
magnetic monopole
resides in that cube.  Of course, since the Dirac strings
themselves
must be unobservable, this theory should respect, as well as
the usual
$SU(N)$ local symmetry, an additional $Z_N$ local symmetry
that deforms
the Dirac strings (without, of course, changing where the
monopoles
are).

The construction of this theory was described by Srednicki
and
Susskind.\refmark{\ss}  In the absence of matter fields, it
has the action
$$
S= -\be \sum_{\rm plaq} {\rm tr}(\eta_P U_P) - \la\sum_{\rm
cubes}
\eta_c
+ c.c.~,
\eqn\Action
$$
where
$$
\eta_c=\prod_{P\in c}\eta_P
\eqn\monopoleB
$$
is the product of the six $\eta_P$'s associated with the
oriented faces
of the cube $c$; in other words, $\eta_c$ is the $Z_N$
magnetic charge
inside $c$.

The extra $Z_N$ local symmetry respected by this theory is
defined on links;
it acts according to
$$
\eqalign{
n:~~&U_l\to e^{2\pi i n/N} U_l~,\cr
&\eta_P\to  e^{-2\pi i n/N}\eta_P~, ~~P\ni l~,\cr}
\eqn\monopoleC
$$
where $n=0,1,2,\dots,N-1$.
This transformation moves the Dirac strings without changing
the
magnetic charge $\eta_c$ or the magnetic flux $\eta_P U_P$
that appears in
the gauge field plaquette action.  (The quantity $U_P$ is
not invariant
under the extra local symmetry, and so is unphysical; it can
be
interpreted as a fictitious magnetic flux that includes a
contribution
from the (unobservable) Dirac string that crosses $P$.)

The coupling constant $\lambda$ controls the strength of the
effects of
virtual monopoles.  For $\lambda <<1$, magnetic charge
fluctuations
occur copiously, but for $\lambda>>1$, magnetic charge
fluctuations are
strongly suppressed.  In the limit $\lambda\to\infty$, the
monopoles
freeze out, and eq.~\Action\ becomes the usual Wilson
action.

Matter fields can be coupled to the gauge theory in the
usual way.  For
example if $\phi_i\in SU(N)$ is defined on sites (labeled by
$i$), we
may define
$$
S_{\rm matter}=-\gamma \sum_{\rm links}
\chi^{(R)}\left((\phi^{-1}U\phi)_l\right)~+c.c.~.
\eqn\monopoleD
$$
This is invariant under the local symmetry eq.~\monopoleC\
only if $Z_N$
is contained in the kernel of the representation $(R)$.  Of
course, this
is just the Dirac quantization condition---the matter fields
must be
chosen so that the string of a monopole is invisible.  We
may introduce
matter fields that are only invariant under  some subgroup
of $Z_N$, but
then we must restrict the $\eta_P$'s to take values in that
subgroup.

\section{Wilson Loop Operator}
Now we want to define a Wilson loop operator $W^{(\nu)}(C)$
that creates
an electric flux tube on the loop $C$.  But the usual
construction
$$
W^{(\nu)}(C)={1\over n_\nu}\chi^{(\nu)}\left(\prod_{l\in C}U_l\right)
\eqn\monopoleE
$$
is not invariant under the local symmetry eq.~\monopoleC\
unless $Z_N$
is contained in the kernel of the representation $(\nu)$.
The operator
$W^{(\nu)}(C)$ is physically sensible, then, only if the
string created
by it has no Aharonov--Bohm interaction with a $Z_N$
monopole.  In fact,
in the $SU(N)$ theory defined by eq.~\Action, this operator
does not
create a stable flux tube at all---even at strong coupling
($\beta<<1$),
glue fluctuations cause the string to
break.

An operator that creates a stable electric flux tube that
{\it does}
have Aharonov--Bohm interactions with monopoles cannot
depend on the
loop $C$ alone; it must also depend on a surface $\Sigma$
that is
bounded by $C$.  After our discussion of the 't Hooft
operator in
Section~5.3, this comes as no surprise.  A (naive) Wilson
loop
operator, in
the presence of monopoles, is a multi-valued object, for it
acquires a
non-trivial Aharonov--Bohm phase when the loop winds around
a magnetic
charge.  To construct a single-valued object, we introduce a
branch
cut on the surface $\Sigma$, so that the operator jumps
discontinuously
when a monopole crosses $\Sigma$.  We augment the naive
Wilson loop
operator, then, by a factor that counts the total magnetic
flux of the
Dirac strings that cross $\Sigma$, obtaining
$$
W^{(\nu)}(C,\Sigma)\equiv
{1\over n_\nu}\chi^{(\nu)}\biggl(
\Bigl( \prod_{l\in C} U_l \Bigr)\Bigl( \prod_{P\in \Sigma}
\eta_P
\Bigr)\biggr)~.
\eqn\modwilson
$$
This operator {\it is} invariant under eq.~\monopoleC, for
any
representation $(\nu)$ of $SU(N)$, and of course it reduces
to
$W^{(\nu)}(C)$  if $(\nu)$ represents $Z_N$ trivially.

We may also consider the degenerate case in which the loop C
shrinks to
a point.  If $(\nu)$ represents $Z_N$ faithfully, then the
operator
$$
G^{(\nu)}(\Sigma)={1\over n_\nu}\chi^{(\nu)}\left(
\prod_{P\in\Sigma}\eta_P\right)
\eqn\magflux
$$
inserts the world sheet of a tube with minimal $Z_N$
electric flux, as a
classical source, on the surface $\Sigma$.  (Of course, if
$(\nu)$ represents $Z_N$ trivially, then the flux tube is
invisible, and the operator is trivial.)  If we think of the
surface
$\Sigma$ as lying in a time slice, then $G(\Sigma)$ has
another
interpretation; it is a magnetic charge
operator that measures the total $Z_N$ magnetic flux through
$\Sigma$.
Obviously, $G(\Sigma)$ is the magnetic analog of the
operator that we
called $F(\Sigma)$ in Section 4.

Note also that if matter is introduced as in eq.~\monopoleD,
an operator
can be constructed that creates a separated
particle--antiparticle
pair; it is
$$
\chi^{(R)}\left(\phi_i^{-1}
\left(\prod_{l\in P_{i,j}}U_l\right)
\phi_j\right)~,
\eqn\lineop
$$
where $P_{i,j}$ is a path from the site $i$ to the site $j$.
This
operator is invariant under the local symmetry, if $(R)$
represents
$Z_N$ trivially.

\section{'t Hooft Loop Operator}
We can also construct an 't Hooft loop operator $B_n(C)$; it
inserts on the world line $C$ a classical monopole source
with $Z_N$ magnetic charge $n= 1, 2,\dots, N-1$.
Alternatively, we may interpret $B_n(C)$, acting in a time
slice, as an operator that creates a loop of magnetic flux
tube on $C$.

The construction of this 't Hooft operator may be carried
out in much the same way as in a gauge theory without
dynamical
monopoles.  We regard $C$ as a closed loop composed of links
of the {\it dual} lattice, and we
chose an arbitrary surface $\Sigma$ (composed of plaquettes
of the dual lattice) whose boundary is $C$.  Dual to the
plaquettes of $\Sigma$ is a set $\Sigma^*$ of plaquettes of
the original lattice.  Now, if there are no dynamical
monopoles, we regard
the $\eta_P$'s as classical (non-fluctuating) variables.
Then, to
evaluate a Green function with an
insertion of $B_n(C)$, we modify the plaquette action on the
plaquettes in $\Sigma^*$, by choosing
$$\eqalign{
&\eta_P = e^{2\pi i n/N}~,
{}~~~P\in \Sigma^*~.\cr
&\eta_P=~~ 1~~~~~~, ~~~P\not\in\Sigma^*~.\cr}
\eqn\tHZN
$$
When there are dynamical monopoles, however, and the
$\eta_P$'s are in
the configuration eq.~\tHZN, the cubes that are dual to the
links of $C$
are frustrated.  Thus, this configuration represents a {\it
dynamical}
monopole propagating on the world line $C$.  Since we want
the operator
$B_n(C)$ to introduce a {\it classical} monopole source, we
should
modify the cube action so that the configuration eq.~\tHZN\
does not
frustrate any cubes.  Thus, we evaluate a Green function
with an
insertion of $B_n(C)$ by changing the cube action according
to
$$
\eta_c\to e^{2\pi i n/N}\eta_c~,
{}~~~c\in C^*~.
\eqn\monoloop
$$
Equivalently, we have
$$
B_n(C)=\prod_{c\in C^*}\exp\left(
\lambda(e^{2 \pi i n/N}\eta_c -\eta_c + c.c)
\right)~.
\eqn\monoloopB
$$
This operator is the magnetic analog of the Wilson loop
operator.

In the weakly--coupled gauge theory without dynamical
monopoles, the 't Hooft loop operator creates a stable
magnetic flux tube, and $\VEV{B_n(C)}$ exhibits area--law
decay.  But when there are dynamical monopoles, $B_n(C)$
always exhibits
perimeter--law decay.  The interpretation is clear.  For any
finite $\lambda$, a $Z_N$ magnetic flux tube is unstable,
for the tube can break via nucleation of a
monopole--antimonopole pair.

Note that the operator defined by eq.~\monoloopB\ makes
sense even if C is an open path rather than a closed loop.
That is, we may define in like fashion an operator
$$
B_n(P_{i,j})=\prod_{c\in P_{i,j}^*}\exp\left(
\lambda(e^{2 \pi i n/N}\eta_c -\eta_c + c.c)
\right)~,
\eqn\monopair
$$
where $i$ and $j$ are sites of the dual lattice, and
$P_{i,j}$ is a path connecting these sites.  This operator
creates a monopole--antimonopole pair, connected by a Dirac
string; it can be used to compute the mass of a dynamical
monopole.
Obviously, it is closely analogous to the operator
eq.~\lineop.

Having now in hand the operator $B_n(C)$ that introduces a
classical $Z_N$ monopole on the world line $C$, and the
operator $G^{(\nu)}(\Sigma)$ that introduces a classical
$Z_N$ electric flux tube on the world sheet $\Sigma$, we are
ready to construct the operator, analogous to
$A^{(\nu)}_n(\Sigma,C)$, that probes the Aharonov--Bohm
interaction between monopoles and electric flux tubes; it is
$$
E^{(\nu)}_n(\Sigma,C)={G^{(\nu)}(\Sigma)~B_n(C)\over
\VEV{G^{(\nu)}(\Sigma)}~\VEV{B_n(C)}}~.
\eqn\Edef
$$
If there is an infinite range Aharonov--Bohm interaction,
the expectation value of this operator will have the
asymptotic behavior
$$
{\rm lim}~\VEV{E^{(\nu)}_n}=
{1\over n_\nu}\chi^{(\nu)}\left(\left(e^{2\pi i n/N}\right)
^{k(\Sigma,C)}\right)~,
\eqn\Elim
$$
where $k(\Sigma, C)$ is the linking number of $\Sigma$ and
$C$.

Finally, we remark that it is straightforward to generalize
the construction in Section~5, and define an operator that
creates a loop of {\it nonabelian} cosmic string, in a
theory that contains dynamical monopoles.  We need only be
cognizant of the change in the plaquette action that occurs
when dynamical monopoles are included; a string that carries
flux $a\in SU(N)$ is now created by
$$
B_a(C,\Sigma,x_0) =  \prod_{P\in\Si^*}
\exp \Bigl\{ \be \Bigl( {\rm tr}(V_{l_P}aV_{l_P}^{-1}\eta_P
U_P) -
{\rm tr}(\eta_P U_P) \Bigr) + c.c.\Bigr\}~,
\eqn\modthooft
$$
where $\Sigma$ is a surface on the dual lattice, bounded by
the loop $C$.  When $C$ shrinks to a point, we obtain the
operator $F_a(\Sigma,x_0)$ that introduces a classical
string source on the closed world sheet $\Sigma$.

\section{Monopole Condensation}
A pure $SU(N)$ gauge theory, without dynamical monopoles, is
confining at strong coupling.  For $\beta<<1$, gauge field
fluctuations are unsuppressed, and the resulting magnetic
disorder gives rise to stable electric flux tubes.  We want
to explore how dynamical magnetic monopoles modify the
physics of this theory.

First, we consider the parameter regime $\beta<<1$ and
$\lambda<<1$, so that virtual monopoles are unsuppressed.
It is easy to anticipate what will happen.  A ``monopole
condensate'' will form, which, in effect, will spontaneously
break the local $Z_N$ symmetry of the theory.  Thus, the
electric flux tube will become the boundary of a domain
wall.  As usual, this domain wall will decay by quantum
tunneling---an electric flux tube will spontaneously
nucleate, and expand, consuming the wall.  Thus, there will
be no stable electric flux tubes, and no infinite range
Aharonov--Bohm interaction between flux tubes and monopoles.

We can check whether this expectation is correct in
strong--coupling perturbation theory.  We proceed by
expanding
$e^{-S_{\rm plaq}}$ in powers of $\beta$ at each plaquette,
and $e^{-S_{\rm cube}}$ in powers of  $\lambda$ at each
cube.  Roughly speaking, the terms that survive when the
$U_l$'s and $\eta_P$'s are summed are ones such that a set
of ``tiled'' cubes forms a closed (three-dimensional)
hypersurface, or else the tiled cubes form an open
hypersurface that is bounded by a (two-dimensional) surface
of tiled plaquettes.  In other words, strong--coupling
perturbation theory can be interpreted as a sum over
histories for
(heavily suppressed) domain walls bounded by electric flux
tubes.

Consider, now, the behavior of $G^{(f)}(\Sigma)$, where
$(f)$ denotes the defining representation of $SU(N)$.  If
$\beta=0$, then the leading contribution to
$\VEV{G^{(f)}(\Sigma)}$, for $\lambda<<1$, is obtained by
tiling the minimal hypersurface that is bounded by $\Sigma$;
thus we have
$$
\VEV{G^{(f)}(\Sigma)}\sim (\lambda)^{{\rm Volume}(\Sigma)}~.
\eqn\wallvol
$$
The interpretation is that $G^{(f)}(\Sigma)$ is the boundary
of a domain wall, where the wall tension is $\sigma\sim -
\epsilon^{-3}~{\rm ln}(1/\lambda)$ ($\epsilon$ is the
lattice spacing).  But for $\beta>0$, this domain wall is
unstable.  When the surface $\Sigma$ is very large, a much
larger contribution to $\VEV{G^{(f)}(\Sigma)}$ is obtained
by tiling the plaquettes of $\Sigma$; this contribution is
$$
\VEV{G^{(f)}(\Sigma)}\sim (\beta/N)^{{\rm Area}(\Sigma)}~.
\eqn\wallarea
$$
(As $\beta\to 0$, the electric flux tube becomes arbitrarily
heavy, and the domain wall is arbitrarily long--lived.)

Accordingly, the operator $W^{(f)}(C,\Sigma)$ does not
create a stable electric flux tube.  The leading behavior of
its expectation value, too, is found by tiling the
plaquettes of $\Sigma$, so that
$$
\VEV{W^{(f)}(C,\Sigma)}\sim
(\beta/N)^{{\rm Area}(\Sigma)}~.
\eqn\modWilsonarea
$$
There is no dependence on the area of the minimal surface
bounded by $C$ (compare eq.~\tensionsimple), signifying that
the dynamical
flux tube created by $W^{(f)}(C,\Sigma)$ has vanishing
tension.

It is obvious that the leading contribution to
$\VEV{G^{(f)}(\Sigma)}$ is unaffected by an insertion of
$B_n(C)$, even if $C$ and $\Sigma$ link.  So we have
$$
{\rm lim}~\VEV{E^{(f)}_n(\Sigma, C)}=1~;
\eqn\trivmaglim
$$
there is no long--range Aharonov--Bohm interaction.

Incidentally, we could have chosen the cube action to be
$$
S_c=-~\lambda (\eta_c)^m + c.c~.
\eqn\chargemmono
$$
Then, in effect, for $\lambda<<1$, charge-$m$ monopoles
condense; this breaks the local symmetry to $Z_M$, where $M$
is the greatest common factor of $N$ and $m$ (compare
Section~4).  In other words, $W^{(\nu)}(C,\Sigma)$ {\it
does} create a stable flux tube, if a source transforming
as the representation $(\nu)$ has no
Aharonov--Bohm interaction with a charge-$m$ monopole, and
if $(\nu)$ represents $Z_M$
nontrivially.

\section{Magnetic Hair}
Now we consider the regime $\beta<<1$ and $\lambda>>1$.  In
this limit, virtual monopoles are heavily suppressed.  We
anticipate that quantum--mechanical magnetic charge
fluctuations will {\it not} wipe out the infinite--range
Aharonov--Bohm interaction between monopoles and electric
flux tubes.

Again, we can check this expectation against perturbation
theory.  Weak--coupling perturbation theory, for
$\lambda>>1$ is an expansion in the number of frustrated
cubes.  The frustrated cubes form closed loops that we may
interpret as the world lines of virtual monopoles.  The
expansion in $\beta$, as before, is an expansion in the
number of tiled plaquettes.  The tiled plaquettes form
closed surfaces that we may interpret as the world sheets of
electric flux tubes.  Perturbation theory, then, is a sum
over  histories for (heavily suppressed) magnetic monopoles
and electric flux tubes.

For example, consider $\VEV{G^{(f)}(\Sigma)}$.  The leading
nontrivial contribution arises from a configuration such
that (in a particular gauge) $\eta_P=e^{\pm 2\pi i /N}$ for
a single plaquette $P$ contained in $\Sigma^*$, while
$\eta_P=1$ for all other plaquettes.  Flipping one plaquette
frustrates four cubes, so this contribution gives
$$
\VEV{G^{(f)}(\Sigma)}\sim 1+\sum_{n=\pm 1}
\left(e^{2\pi i n/N}-1\right)
\left(\exp\left[-2\lambda\left(
1-\cos\left({2\pi\over N}\right)\right)
\right]\right)^4~.
\eqn\leadG
$$
Summing disconnected contributions causes the result to
exponentiate; we find
$$
\VEV{G^{(f)}(\Sigma)}\sim \exp\left(
-\kappa^{(\rm ren)}{\rm Area}(\Sigma)\right)~,
\eqn\expores
$$
where
$$
\epsilon^2 \kappa^{(\rm ren)}\sim
2\left(1-\cos(2\pi/N)\right)\exp\left(
-8\lambda\left[1-\cos\left({2\pi\over N}\right)\right]
\right)
\eqn\electenren
$$
(and $\epsilon$ is the lattice spacing).  The interpretation
is that, because of the Aharonov--Bohm interaction between
monopoles and flux tubes, inserting $G^{(f)}(\Sigma)$
modifies the contribution to the vacuum energy due to
virtual
monopole pairs that wind around $\Sigma$, resulting in a
renormalization of the tension of the classical string
source.

When we compute $\VEV{W^{(f)}(C,\Sigma)}$, a similar
renormalization of the tension of the classical string on
$\Sigma$ occurs.  But in addition, for $\beta<<1$, the
configurations that contribute have a surface of tiled
plaquettes bounded by $C$.  The leading behavior, then, is
$$
\VEV{W^{(f)}(C,\Sigma)}\sim (\beta/N)^{{\rm Area}(C)}
{}~\exp\left(
-\kappa^{(\rm ren)}{\rm Area}(\Sigma)\right)~.
\eqn\monoWilbeh
$$
We conclude that $W^{(f)}(C,\Sigma)$ creates a stable
electric flux tube with string tension
$$
\epsilon^2\kappa^{(\rm dyn)}\sim
-{\rm ln}(\beta/N)~.
\eqn\sttension
$$

When the operator $B_n(C)$ is inserted, it tends to
frustrate the cubes in $C^*$.  But, as we already noted in
Section~10.3,
frustrated cubes can be
avoided if the $\eta_P$'s assume a suitable configuration.
We may choose an arbitrary surface ${\rm T}$ on the dual
lattice whose boundary is $C$.  Dual to the plaquettes of
${\rm T}$ is a set of plaquettes ${\rm T}^*$ of the original
lattice.  The desired configuration (in a particular gauge)
is
$$
\eqalign{
&\eta_P=e^{2\pi i n/N}~,~~~P\in{\rm T}^*~,\cr
&\eta_P=~~1~~~~~~,~~~P\not\in {\rm T}^*~.\cr}
\eqn\nofrust
$$
(The local symmetry transformation eq.~\monopoleC\ deforms
the surface $T$, but leaves its boundary intact.)

By summing over gauge field fluctuations about the
configuration
eq.~\nofrust, we find the leading behavior
$$
\VEV{B_n(C)}\sim \exp\left(-M_n^{(\rm ren)}~{\rm Perimeter}
(C)\right)~,
\eqn\maghairA
$$
where
$$
\epsilon M_n^{(\rm ren)}\sim 2N^2(\beta/N)^6
\left(1-\cos\left({2\pi n\over N}\right)\right)
\eqn\maghairB
$$
is the renormalization of the mass of the classical monopole
source.
This renormalization is associated with virtual electric
flux tubes
whose world sheets link the world line of the monopole, and
arises
because of the Aharonov--Bohm interaction between  monopole
and flux
tube.

When the operator $B_n(P_{i,j})$ is inserted, a line of
frustrated cubes
connecting $i$ and $j$ cannot be avoided, and so we find the
leading
behavior
$$
\VEV{B_n(P_{i,j})}\sim\exp\left(-M_n^{(\rm dyn)}~{\rm
Distance}(i,j)
\right)~\exp\left(-M_n^{(\rm ren)}~{\rm Length}(P)\right)~,
\eqn\maghairC
$$
where
$$
\epsilon M_n^{(\rm dyn)}\sim 2\lambda
\left(1-\cos\left({2\pi n\over N}\right)\right)~.
\eqn\maghairD
$$
Evidently, for $\lambda>>1$ and $\beta<<1$, we have
$M_n^{(\rm dyn)}>>
M_n^{(\rm ren)}$.  Indeed,  $\VEV{B_n(C)}$ is dominated by
small fluctuations about the configuration eq.~\nofrust\ for
precisely
this reason---the renormalization of the classical source is
much less
costly than screening the source with dynamical monopoles.

With the $\eta_P$'s in the configuration eq.~\nofrust, the
operator $G^{(f)}(\Sigma)$  assumes the value $\exp(2\pi i n
k/N)$, where $k$ is the linking number of $\Sigma$ and $C$.
Furthermore, except on the loop $C$, this configuration is
locally equivalent to the trivial configuration with
$\eta_P=1$ everywhere.
Thus, as we expand in the small
fluctuations about eq.~\nofrust, we find, to each order of
the expansion,
$$
{\rm lim}~\VEV{E^{(f)}_n}=\left(e^{2\pi i n/N}\right)
^{k(\Sigma,C)}~.
\eqn\monoAByes
$$
We see that, at least to each order of perturbation theory,
our expectation is confirmed.  In a confining theory that
contains weakly coupled dynamical magnetic monopoles, there
is an infinite--range Aharonov--Bohm interaction between
monopoles and electric flux tubes.

\bigskip
We thank Hoi-Kwong Lo, and, especially, Martin
Bucher, for helpful discussions. This work was supported
by DOE grant DE-AC03-81-ER40050, and by NSF
grants NSF-PHY-87-14654, NSF-PHY-90-21984
and NSF-PHY-89-04035. JMR and
MGA wish to thank the Aspen Center for Physics and,
especially,
the Caltech Physics Department, for their hospitality
during portions of this work.

\def\Z{{Z}}

\let\be=\beta

\let\de=\delta

\let\ep=\epsilon

\let\ka=\kappa
\let\la=\lambda

\let\Si=\Sigma
\let\th=\theta

\let\om=\omega
\let\Om=\Omega
\let\p=\partial
\let\<=\langle
\let\>=\rangle

\def\Re{{\rm Re}\,}

\def\comment#1{ \hbox{Comment suppressed here.} }

\tolerance=10000
\appendix
In this appendix we consider in more detail some of the
lattice perturbation theory
calculations mentioned in the body of the text.
Specifically, we calculate, in a pure gauge theory, the
behavior of
$\< F_a(\Si,x_0)\>$ and $\< B_a(\Si,C,x_0)\>$
in leading order in both strong and weak coupling
perturbation theory.
We demonstrate that $\< F_a(\Si,x_0)\>$ exhibits an area law
decay
in both limits, and that only in the weak coupling limit
does
$\< B_a(\Si,C,x_0)\>$ create a dynamical string on $C$.
We also consider the problem that arises when one attempts
to use the untraced Wilson loop operator eq.~\Udef\  to
construct
an Aharonov-Bohm order parameter.

\section{Pure gauge theory: strong coupling}
We start with
the calculation of $\< F_a(\Si,x_0)\>$ and $\<
B_a(\Si,C,x_0)\>$ in the
strong coupling regime ($\be << 1$) of the pure gauge theory
defined by the plaquette action eq.~\plq, namely
$$
S^{(R)}_{{\rm gauge},P} = -\be\chi^{(R)}(U_P)+ c.c.
\eqn\plqagain
$$
For definiteness we will assume that the representation
$(R)$ that defines
the theory is irreducible, and that it satisfies the
constraints that
$R\otimes R$ does not contain the trivial representation,
while $R\otimes
R^*$ contains it exactly once.

The expectation value $\< F_a(\Si,x_0)\>$ is given by
$$
\< F_a(\Si,x_0)\>={ \<\< F_a(\Si,x_0)\>\>\over \<\< 1 \>\>
},
\eqn\expect
$$
where the unnormalized expectation value of an operator $X$
is
defined by
$$
\<\< X \>\>=\prod_{\< ij\>}\int dU_{\< ij\>}
\Bigl( X \prod_{P}\exp(-S_P) \Bigr).
\eqn\unnorm
$$
To find the renormalized
string tension we need to calculate the leading behavior of
both
$\<\< F_a(\Si,x_0)\>\>$ and $\<\< 1 \>\>$.
The perturbation expansion in the strong coupling regime is
of the form
of a sum over closed surfaces, in which surfaces of greater
area are
suppressed by powers of $\be$ compared to smaller ones.
Formally, the strong coupling expansion proceeds by
performing a
character expansion on the exponentiated Wilson action,
$$
\exp\bigl(\be(\chi^{(R)}(U_P)+\chi^{(R^*)}(U_P))\bigr)
=N(\be)\Bigl(1 + \sum_{(\nu)\neq 1}
C^{(\nu)}(\be)\chi^{(\nu)}(U_P)\Bigr).
\eqn\charx
$$
The link integrations then select out closed surfaces $S$,
each formed by ``tiling''
$S$ with factors of $\chi^{(\nu)}(U_P)$ for each plaquette
$P\in S$.
The leading non-trivial contribution will be from the
smallest
surfaces, which furthermore are tiled with characters whose
associated
factors of $C^{(\nu)}(\be)$ have the lowest non-trivial
dependence
on $\be$. The factors $C^{(\nu)}(\be)$ are found by
multiplying eq.~\charx\  by $\chi^{(\nu^*)}(U_P)$, summing
$U_P$ over
the group, and using the character orthogonality relations
along with
the assumed properties of the representation $(R)$, giving
$$
C^{(R)}(\be)=1+\be +O(\be^2),~~~~C^{(R^*)}(\be)=1+\be
+O(\be^2),
\eqn\charco
$$
with all other irreducible representations acquiring their
first
non-trivial $\be$ dependence at $O(\be^2)$. Therefore the
leading
contribution will come from surfaces tiled with characters
in
the representations $(R)$ and $(R^*)$.

\FIG\strongF{The leading contribution to
$\VEV{F_a(\Sigma,x_0)}$, in the strong--coupling limit of
the pure gauge theory, arises from tiling a cube (shaded)
that contains two plaquettes of $\Sigma^*$.}
In the evaluation of $\< F_a \>$
we must, in addition, identify the leading contribution
that is sensitive to the presence of $F_a$, and hence does
not cancel
between $\<\< F_a \>\>$ and $\<\< 1 \>\>$. All closed
surfaces that do not intersect $\Si$ are unaffected by the
presence
of $F_a$. Therefore we factorize both $\<\< F_a \>\>$ and
$\<\< 1 \>\>$
in eq.~\expect\  into contributions from surfaces ``on-
$\Si$'' and
``off-$\Si$'', the off-$\Si$ dependence cancelling between
them.
The dominant on-$\Si$ contribution is shown in
Fig.~\strongF, where
all six faces are tiled with factors of either $\chi^{(R)}$,
or
$\chi^{(R^*)}$. This results in a contribution to $\<\< F_a
\>\>$
from Fig.~\strongF of
$$
2({\be / n_R})^6 \bigl|\chi^{(R)}(a)\bigr|^2,
\eqn\interm
$$
where $n_R$ is the dimension of the representation $(R)$.
The factor of $(n_R)^{-6}$ comes from the six non-trivial
link
integrations, and the factor of $2$ from the two equal
contributions
of either $\chi^{(R)}$ or $\chi^{(R^*)}$.
Since the number of possible positions of this elementary
cube
on the surface $\Si$ is given by the area $A(\Si)$ of $\Si$
(measured in terms of the size of the elementary plaquette
$\ep^2$),
to get the contribution of a single elementary cube located
anywhere on $\Si$ we should further multiply \interm\ by
$A(\Si)$.

The area law decay of $\< F_a \>$ arises from considering an
arbitrary number of such elementary tiled cubes on the
surface $\Si$,
rather than just a single contribution.
In the ``dilute gas'' approximation where we can ignore
correlations
between the various elementary cubes, this leads to an
exponentiation of the above result with a overall
factor of $A(\Si)$.  Finally, the ``on $\Si$''
contributions to $\<\< 1 \>\>$
lead to a similar exponential dependence on $A(\Si)$,
except that the group element $a$ is replaced by the
identity element. Putting these together yields the result
$$
\< F_a(\Si,x_0)\>= \exp\left(2\bigl(\be/
n_R\bigr)^6\Bigl(|\chi^{(R)}(a)|^2 -
n_R^2\Bigr)A(\Si)\right).
\eqn\appfdecay
$$
In other words, the renormalized classical string tension
defined in eq.~\Fdecay\  is given by
$$
\epsilon^2\ka_a^{(ren)}\sim 2
\left({\be\over n_R}\right)^6
\left(n_R^2-\left|\chi^{(R)}(a)\right|^2\right)
\eqn\renstrten
$$
to lowest order in $\be$.

The calculation of $\<B_a(\Si,C,x_0)\>$ proceeds in a very
similar way;
the only question concerns the behavior at the boundary
curve $C$.
Here there exist contributions to $\<B_a(\Si,C,x_0)\>$
similar to
those of Fig.~\strongF except only one plaquette $P\in
\Si^*$
is contained in the cube. This leads to an additional decay
of
$\<B_a(\Si,C,x_0)\>$ depending on the length $P(C)$ of the
perimeter
(eq.~\perimeter),
$$
\<B_a(\Si,C,x_0)\>\sim \exp\left(-\ka_a^{(ren)}A(\Si)\right)
\exp\left(-m_a^{(ren)}P(C)\right),
\eqn\appbdecay
$$
with
$$
\ep m_a^{(ren)} \sim 2
\left({\be\over n_R}\right)^6
\left(n_R^2-n_R\Re\chi^{(R)}(a)\right).
\eqn\appmren
$$

Since the configurations that dominate the expectation
values of
$F_a$ and $B_a$ do not extend deep into the volume enclosed
by
$\Si$, we find that there are no stable dynamical strings
(or vortices)
in this regime, and therefore that there is no long-distance
Aharonov-Bohm interaction between flux tubes and charges.

\section{Weak coupling}

\FIG\weakF{The leading contribution to
$\VEV{F_a(\Sigma,x_0)}$,
in the weak--coupling limit of the pure gauge theory.  The
cube shown contains two plaquettes that are in $\Sigma^*$,
and
the link variables assume the indicated values.  (Unmarked
links
have the value $U_l=1$.)  There are altogether nine excited
plaquettes---eight (unshaded) plaquettes with $U_P=b$ and
one
(shaded) plaquette with $U_P=aba^{-1}b^{-1}$.}
This is the regime in which we expect to find
$A_a^{(\nu)}(\Sigma,x_0;C)$ displaying a non-trivial
Aharonov-Bohm
interaction. To start we need to discover which
configurations
dominate the behavior of $\< F_a(\Sigma,x_0)\>$.
When $\be >>1$, frustrated plaquettes
(ones with non-minimal gauge action) are highly suppressed.
However, as described in Section~6.2, an insertion of
$F_a(\Si,x_0)$
tends to frustrate the plaquettes in $\Si^*$, unless the
link variables
assume a suitable configuration. The configuration that
leaves
no frustrated plaquettes, and thus acts as the ``ground
state''
in the presence of $F_a$ is illustrated in Fig.~9. It
consists
of a ``forest'' of links inside $\Si$ that have value
$U_l=a$. Specifically we choose a three-dimensional
hypersurface
$\Om$ made up of cubes of the dual lattice that has boundary
$\Si$. Dual to these cubes is a set of links $\Om^*$ on the
original
lattice. The configuration with no excited plaquettes is
then
given by eq.~\config, namely $U_l=a$ for $l\in \Om^*$, and
$U_l=e$
otherwise. The area law decay of $\< F_a(\Si,x_0)\>$
in the weak coupling regime arises from fluctuations of the
link
variables around this state, which are sensitive to the
presence
of the non-trivial background.

The leading such contribution to the decay of $\<
F_a(\Si,x_0)\>$
in the pure gauge theory is shown in Fig.~\weakF a. (For
simplicity
we discuss the situation for vortices in three Euclidean
dimensions.)
It consists of a configuration that has nine excited
plaquettes,
and is constructed by considering two
neighboring plaquettes in $\Si^*$. These plaquettes are
connected
with four links with $U_l=b$ to make a cube, with $b$ summed
over the group. In this configuration,
there are eight excited plaquettes with
plaquette action proportional to $\be\Re\chi(b)$ (the
unshaded
plaquettes in Fig.~\weakF b). But, in addition, there is an
excited plaquette (if $a$ and $b$ do not commute)
with action proportional to
$\be\Re\chi(aba^{-1}b^{-1})$ (the shaded plaquette in
Fig.~\weakF b).
This extra excitation distinguishes the contributions of
Fig.~\weakF\ to
$\<\< F_a(\Si,x_0)\>\>$ and $\<\< 1 \>\>$.
Changes of variable can move the ninth excited plaquette
around, but cannot get rid of it.

Of course, the process illustrated in Fig.~\weakF\ is only
one
of a set of similar processes where instead of summing
over the same value $b$ for each of the four connecting
links,
we sum independently over their values $b_i$. These differ
from the
contribution of Fig.~\weakF\ in two ways: there will either
be a greater number of excited plaquettes leading to a
subdominant contribution, or, if some of the $b_i$ are taken
to be the
indentity, changes of variable on the remaining,
independent, $b_i$ show
that the $a$-links have no physical effect. This is why, for
instance,
we do not consider the less suppressed, and apparently non-
trivial,
contribution arising from the excitation of just two joining
links.

We can understand this in physical terms. The contribution
to
$\< F_a(\Si,x_0)\>$ reflects how the non-trivial boundary
condition
introduced by the vortex (or string)
affects the quantum fluctuations of the gauge
fields. To see the effect, we must consider processes in
which virtual
gauge field excitations (with gauge quantum numbers that
Aharonov-Bohm
scatter off the inserted string of flux $a$) propagate
around the string.
The process illustrated in Fig.~\weakF\ is the leading one
of this type.

The end result of these considerations is a renormalization
of
the classical string tension (or vortex mass in three
dimensions)
leading to an area law (respectively, perimeter law) decay
of
$\< F_a(\Si,x_0)\>$ as in eq.~\Fdecay. An estimate of the
renormalized
vortex mass in the weak-coupling regime arising from
Fig.~\weakF\ is
$$\eqalign{
\ep m_a^{(ren)}\sim
{1 \over n_G}\sum_{b\neq 1}\biggl( \exp&\bigl\{ (-
16\be)[n_R-\Re\chi^{(R)}(b)]
   \bigr\}\times \cr
& \Bigl(1- \exp\bigl\{ (-2\be)[n_R - \Re
   \chi^{(R)}(bab^{-1}a^{-1})]\bigr\}\Bigr) \biggr),}
\eqn\weakrenvormass
$$
where we require more information about the group $G$ to
explicitly evaluate the sum.

Also, note that we have found area
law decay of $\< F_a(\Sigma,x_0)\>$ in both regimes of
the pure non-abelian gauge theory described by
eq.~\plqagain\
(when $a$ is not in the center of $G$).
This differs from the situation in
a pure abelian gauge theory where a change of variables
shows that $\< F_a(\Sigma,x_0)\>\equiv 1$.\refmark{\pk}
This is to be expected since the pure non-abelian gauge
theory
contains excitations that, via the Aharonov-Bohm
effect, are sensitive to the introduction of flux on $\Si$.

To see that there exist stable dynamical strings in this
regime
of the pure gauge theory, we must consider
$\< B_a(\Si,C,x_0) \>$, rather than $\< F_a(\Sigma,x_0)\>$.
Since the surface $\Si$ now has a boundary, the three
dimensional
hypersurface $\Om$ must also end on a two dimensional
surface $S$.
This surface is dual to a set of plaquettes $S^*$ on the
original
lattice that must be frustrated, since only one of their
links is
contained within $\Om^*$. Clearly the dominant configuration
is the one in which $S$ is the minimal area surface with
boundary $C$.
Each such plaquette is suppressed relative to the
corresponding
vacuum configuration that dominates $\<\< 1\>\>$ by an
amount
$$
\exp\bigl(2\be(\Re\chi^{(R)}(a) - n_R)\bigr).
\eqn\appfrust
$$
In other words $\< B_a(\Si,C,x_0) \>$ behaves as in
eq.~\tensionsimple
$$
\< B_a(\Si,C,x_0) \>=\exp\bigl(-\ka_a^{({\rm ren})}
A(\Si)\bigr)
\exp\bigl(-\ka_a^{({\rm dyn})} A(C)\bigr),
\eqn\appbdecay
$$
where $A(C)$ is the area of the minimal surface $S$, and
$\ka_a^{({\rm dyn})}$ is the dynamical string tension of the
stable
string with flux in the $a$ conjugacy class,
$$
\ep^2\ka_a^{({\rm dyn})}\sim
2\be\left(n_R - \Re\chi^{(R)}(a)\right).
\eqn\appdynst
$$

In some situations it is possible that $B_a(\Si,C,x_0)$
creates not just one
dynamical string, of flux $a$, on $C$, but instead creates
a number of strings of varying fluxes $a_1,\dots,a_k$,
where $a_1a_2\dots a_k=a$. The corresponding link
configuration
would be a ``forest state'' of links of value
$U_l=a$ that terminates gradually in a number of steps
rather than dropping to the indentity across a single
plaquette.
This is the more favorable configuration if the total area
law
suppression of the multiple strings is less than
that for the single string. This translates into the
condition (in the pure gauge case, and ignoring perimeter
corrections that vanish in the infinite size limit)
$$
(k-1)n_R + \Re\chi^{(R)}(a) -
\sum_{i=1}^{k}\,\Re\chi^{(R)}(a_i) <0~.
\eqn\lesssupp
$$
As an example of this, consider the expectation value, in a
pure
$Z_6$ gauge theory, of the operator $B_3(\Si,C,x_0)$ that
attempts
to create a dynamical string with flux $a=\om^3$ on $C$
(where
$\om=\exp(2\pi i/6)$). The dominant process in this
case is not the creation of an $\om^3$ string, but instead
that
of three $\om$ strings, since, substituting into
eq.~\lesssupp, we find $2+(-1)-3(1/2)<0$.

We now turn to the order parameter
$A_a^{(\nu)}(\Sigma,x_0;C)$, which is defined (eq.~\NAA) in
terms
of $F_a(\Si)$ and the Wilson loop operator
$W^{(\nu)}(C)$. $F_a(\Si)$ has been described above, so we
will
now discuss the Wilson loop operator, and demonstrate
eq.~\NAlim.

First, let us consider the untraced Wilson operator
$U^{(\nu)}(C,x_0)$ (eq.~\Udef).
It is easy to calculate the leading behavior of the
expectation value of $U^{(\nu)}(C,x_0)$
at weak coupling: as above we factorize the contributions
from
links on-$C$ and off-$C$, and cancel the off-$C$
contributions,
$$
{ \<\<U^{(\nu)}(C,x_0)\>\> \over \<\<1\>\> } =
{ \sum_{U_1}\cdots \sum_{U_L} D^{(\nu)}(U_1\cdots U_L)
\prod_{i=1}^L \exp\Bigl(4D\be\Re\chi^{(R)}(U_i) \Bigr)
\over
\sum_{U_1}\cdots \sum_{U_L}
\prod_{i=1}^L \exp\Bigl(4D\be\Re\chi^{(R)}(U_i)\Bigr) }~,
\eqn\expu
$$
where we work in $D+1$ Euclidean dimensions, and
$U_i$ for $i=1\ldots L$ are the links in $C$.  (In
eq.~\expu,
we sum over all values of the link variables contained in
$C$,
but keep all other link variables fixed at $U_l=1$.)
Because we have not traced the Wilson loop the multiple
summations
factorize. We then use
$$
{1\over n_G} \sum_{g\in G} D^{(\nu^*)}(g)\chi^{(\mu)}(g)
= {1\over n_\nu} \de^{\mu,\nu} I^{(\nu)},
\eqn\appident
$$
where $I^{(\nu)}$ is the identity matrix. Along with
eq.~\charx\
this implies that
$$
\< U^{(\nu)}(C,x_0) \> = \left( {C^{(\nu^*)}(2D\be) \over
n_\nu}
\right)^L I^{(\nu)}.
\eqn\appuntrace
$$
If we repeat the calculation in the background created by a
string insertion operator $F_a(\Si,x_0)$ then the only
difference
is that for each time $C$ links $\Si$ there is a link in $C$
that is multiplied by $a$. Neither $F_a(\Si,x_0)$ nor
$U^{(\nu)}(C,x_0)$
is gauge-invariant, but a gauge transformation conjugates
both
$a$ and $U^{(\nu)}$ by the value of the transformation at
$x_0$,
so these conjugations cancel, and the gauge-invariant result
is
that $I^{(\nu)}$ is replaced by
$D^{(\nu)}(a^{k(\Si,C)})$, where $k(\Si,C)$ is the linking
number of the two curves. This proves eq.~\untraced, and by
taking
a trace it proves eq.~\NAlim.

However there is a major caveat that limits the usefulness
of
the untraced Wilson loop: there are corrections to
eq.~\untraced\
that do not go to zero as $C$ and $\Si$ become large and
well separated.
Recall that $F_a(\Si,x_0)$ corresponds to
a thought experiment in which a specific group element $a$
(not just a conjugacy class) was associated with a string,
using
an arbitrary basis at a point $x_0$. As described in
Section~5.2
this means that in defining $F_a(\Si,x_0)$
we give each plaquette in $\Si^*$ a long ``tail''
of links that stretches back to $x_0$.
The problem manifests itself (in a particular gauge)
when we take into account
configurations in which some (but not all) of the links
coming out of
$x_0$ are excited. In these configurations the inserted
flux may be conjugated, but not the measured group element
(or vice versa).

Concretely, the leading contribution, in weak
coupling perturbation theory, that causes problems for the
untraced Wilson loop is simply the excitation of a
single link $U_l=b$ on the path that connects $x_0$ to
$\Si$.
In the three dimensional case this leads to the excitation
of four plaquettes, and is thus suppressed relative
to the ground state configuration by a factor of
$\exp(8\be(\Re\chi^{(R)}(b) - n_R))$. However, the action
of these configurations, though large at weak coupling,
is completely independent of the size and
separation of $\Si$ and $C$, and so remains finite
as $\Si$ and $C$ become large and far apart.
Their effect on the untraced order parameter is to
conjugate the inserted flux (and thus the flux measured by
$U^{(\nu)}(C,x_0)$). Therefore, there are corrections to
eq.~\untraced\
that occur at finite order in weak coupling perturbation
theory that
render the measured flux uncertain up to conjugation.

The physical interpretation of these
configurations is that the path connecting $x_0$ to $\Si$
is linked by a virtual vortex-antivortex pair of flux $b$.
In four dimensions there exist analogous processes where a
virtual cosmic string links the connecting path.
These effects do not correct the leading behavior of the
ABOP
given by eq.~\NAlim, because
the traced Wilson loop is gauge-invariant,
and so is not sensitive to conjugation of the inserted flux.

\refout
\figout

\bye